\documentclass[fleqn,usenatbib]{mnras}

\usepackage{newtxtext,newtxmath}

\usepackage[T1]{fontenc}

\DeclareRobustCommand{\VAN}[3]{#2}
\let\VANthebibliography\thebibliography
\def\thebibliography{\DeclareRobustCommand{\VAN}[3]{##3}\VANthebibliography}

\usepackage{graphicx}	
\usepackage{amsmath}	
\usepackage{xcolor}
\usepackage{ulem}

\newcommand{\Msun}{\,\mathrm{M}_\odot}
\newcommand{\feh}{\mathrm{[Fe/H]}}
\newcommand{\ttid}{t_{\mathrm{tid}}}
\newcommand{\Mc}{M_{\mathrm{c}}}

\newcommand{\Mh}{M_{\mathrm{h}}}

\newcommand{\lambdam}{\lambda_{\mathrm{m}}}

\newcommand{\kms}{\mathrm{\,km\,s^{-1}}}

\title[GC kinematics]{Modeling the kinematics of globular cluster systems}

\author[Y. Chen and O. Y. Gnedin]{Yingtian Chen\thanks{E-mail: ybchen@umich.edu} \href{https://orcid.org/0000-0002-5970-2563}{\includegraphics[scale=0.3]{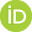}} and
Oleg Y. Gnedin \href{https://orcid.org/0000-0001-9852-9954}{\includegraphics[scale=0.3]{figures/orcid.png}}
\\
Department of Astronomy, University of Michigan, Ann Arbor, MI 48109, USA
}

\date{Accepted XXX. Received YYY; in original form ZZZ}

\pubyear{2022}

\begin{document}
\label{firstpage}
\pagerange{\pageref{firstpage}--\pageref{lastpage}}
\maketitle

\begin{abstract}
Globular clusters (GCs) are old massive star clusters that serve as `fossils' of galaxy formation. The advent of \textit{Gaia} observatory has enabled detailed kinematics studies of the Galactic GCs and revolutionized our understanding of the connections between GC properties and galaxy assembly. However, lack of kinematic measurements of extragalactic GCs limits the sample size of GC systems that we can fully study. In this work, we present a model for GC formation and evolution, which includes positional and kinematic information of individual GCs by assigning them to particles in the Illustris TNG50-1 simulation based on age and location. We calibrate the three adjustable model parameters using observed properties of the Galactic and extragalactic GC systems, including the distributions of position, systemic velocity, velocity dispersion, anisotropy parameter, orbital actions, and metallicities. We also analyze the properties of GCs from different origins. In outer galaxy, \textit{ex-situ} clusters are more dominant than the clusters formed \textit{in-situ}. This leads to the GC metallicities decreasing outwards due to the increasing abundance of accreted, metal-poor clusters. We also find the \textit{ex-situ} GCs to have greater velocity dispersions and orbital actions, in agreement with their accretion origin.
\end{abstract}



\section{Introduction}
\label{sec:introduction}

Globular cluster (GC) systems are widely considered as `fossils' of galaxy formation and evolution \citep{searle_compositions_1978,harris_globular_1991}. A typical GC consists of $10^5 - 10^6$ stars, which are formed within a relatively short time interval less than a few $\rm Myr$ \citep{lada_embedded_2003,mac_low_control_2004,hartmann_rapid_2012}. Therefore, a GC can be regarded as a snapshot of the stellar population that records the local physical and chemical environment at the time of formation. This has motivated multi-wavelength observations of galactic and extragalactic GC systems and helped the development of theories of galaxy formation and evolution. Based on the photometric studies of red giants in galactic GCs, \citet{searle_compositions_1978} discovered a significant age spread of GCs at different regions of the Milky Way (MW). In addition to the traditional photometric studies of GC systems, kinematic studies of MW GCs \citep{cudworth_space_1993,dinescu_space_1999} further revealed the hierarchical nature of the formation of the MW GC system and the MW galaxy itself \citep[see a recent review by][]{helmi_streams_2020}. Recently, the launch of the \textit{Gaia} space observatory \citep{gaia_collaboration_gaia_2016,gaia_collaboration_gaia_2018,gaia_collaboration_gaia_2021} has enhanced our knowledge of the kinematics and structure of the MW GC system in the 6-dimensional phase space, which has improved our understanding of how GCs are formed and brought to the MW \citep{massari_power_2017,helmi_merger_2018,koppelman_characterization_2019,massari_origin_2019}.

It is much harder to measure the 3D velocities of extragalactic GC systems, which limits the sample size of GC systems that we can fully study observationally. However, a larger sample set is crucial for understanding the origin of GC systems, as the statistical noise of a small sample set can conceal connections between physical properties. Therefore, an alternate approach is gaining polarity: numerical modeling. High-resolution simulations that can fully resolve giant molecular clouds (GMCs), which are commonly believed to be the cradles of star clusters, have revealed likely scenarios of cluster formation and evolution \citep[see, e.g.,][]{howard_universal_2018,grudic_nature_2019,li_disruption_2019,ma_self-consistent_2020,chen_effects_2021,grudic_model_2021}. These simulations, with mass resolution down to sub-$\Msun$ scale, have provided fruitful information on the kinematic and structural properties of star clusters. However, these simulations mostly focus on isolated GMCs over a relatively short time interval of several $\rm Myr$. It is therefore hard to include the effects of the cosmological environment on cluster formation and evolution with GMC-scale simulations. To study the origin of GC systems in the cosmological context, one can either run galaxy formation simulations with detailed implementation of cluster formation \citep[e.g.,][]{li_star_2018,li_star_2019}, or post-process outputs of existing simulations with analytical models. The former can better track the evolution of clusters but is computationally expensive. Consequently, the sample size of GC systems in these simulations is usually small, leading to difficulty in characterizing scaling relations between cluster properties.

In contrast, post-processing methods enable us to study the effects of various physical parameters on the origin and disruption of GC systems in a simplified but efficient way. There are several implementations of the post-processing approach, including the MOSAICS model \citep{kruijssen_photometric_2008,kruijssen_evolution_2009,kruijssen_modelling_2011}, our previous models \citep{muratov_modeling_2010,li_modeling_2014,choksi_formation_2018,choksi_formation_2019,choksi_origins_2019}, and other implementations \citep[e.g.,][]{renaud_origin_2017,creasey_globular_2019,phipps_first_2020,halbesma_globular_2020}.  These models need to carefully handle two issues: how to model GC formation and how to model GC evolution. The formation rate of GCs can be calculated from either the local gaseous environment or the global properties of host galaxies. Once a population of GCs is formed, the model determines the mass of each GC following some initial mass function. Next, the model needs to track the mass loss of each GC due to stellar evolution and dynamical disruption. The interplay of formation and evolution shapes the present-day properties of a GC system. By applying the GC formation and evolution model to existing cosmological simulations, we can greatly increase the sample size of GC systems without requiring too many computational resources. Also, the post-processing method is more robust since it separates the prescriptions for GC formation and evolution from the still-uncertain sub-grid models of star formation and feedback in cosmological simulations.

\citet{choksi_origins_2019} developed such an analytical model which can be applied to dark matter-only simulations. By linking GC formation to the mass growth of the host halo, this model has successfully reproduced important observed scaling relations of GC systems with a broad range of galaxy mass. However, the model did not incorporate modeling of the spatial distribution or kinematics of GC systems. This motivates us to develop a new model that links GCs to particles in an underlying simulation. We still keep the dependence on the adopted simulations to a minimal level to preserve the clarity of our model. This new model enables a more detailed description of the dynamical disruption of GCs. In addition, it can be extended to study the assembly of extragalactic GC systems by accretion of satellite galaxies. As we show below, compared with similar works that can also model the spatial distribution of GCs in cosmological simulations \citep[e.g.,][]{ramos-almendares_simulating_2020,trujillo-gomez_kinematics_2021}, our model produces a better match to the observed spatial distribution and kinematics of the MW GCs.

The paper is organized as follows. In Sec.~\ref{sec:methods} we outline the modeling of GC formation and evolution and introduce the sampling of cluster particles from a cosmological simulation. Then we describe the calibration of model parameters using observed scaling relations in Sec.~\ref{sec:model_calibration}. In Sec.~\ref{sec:comparison_with_observations}, we compare the spatial distribution and kinematics of model GC systems with observations. Next, we investigate the properties of GCs from different origins in Sec.~\ref{sec:gcs_from_different_origins}. We discuss how different modeling of GC disruption influences the radial distribution of GC systems, GC mass function, and comparison with other studies in Sec.~\ref{sec:discussion}. Finally, we summarize our results in Sec.~\ref{sec:summary}.

\section{Model for Cluster Formation and Evolution}
\label{sec:methods}

In this work, we propose a new GC formation and evolution model, which post-processes snapshots of cosmological simulations and produces a catalogue of GCs without rerunning the simulation. This catalogue provides such properties as the mass, metallicity, age, position, and velocity of each surviving GC at the present time. To follow the spatial and kinematic distributions of GCs, we use snapshots of the hydrodynamic simulation Illustris TNG50-1 \citep[][hereafter TNG50]{nelson_first_2019,pillepich_first_2019,nelson_illustristng_2021}. A detailed description of the model follows in Sec.~\ref{sec:modeling_cluster_formation_and_evolution}.

\subsection{Background cosmological simulation}
\label{sec:cosmological_simulation}

Our model can be applied to any cosmological simulation, whether purely collisionless or with modeling of gas dynamics and star formation. In this work, we base the model on the simulation suite TNG50, which is performed with the moving mesh, finite-volume hydrodynamic code \textsc{arepo} \citep{springel_e_2010}. TNG50 adopts a flat $\Lambda$CDM universe with cosmological parameters given by the \citet{planck_collaboration_planck_2016}: $\Omega_{\rm b}=0.0486$, $\Omega_{\rm m}=0.3089$, $\Omega_\Lambda=0.6911$, $h=0.6774$, $\sigma_8=0.8159$, and $n_{\rm s}=0.9667$. For consistency, we adopt the same cosmology in our model. TNG50 is initiated with $2160^3$ dark matter particles and the same number of gas cells within a $51.7\ {\rm Mpc}$ comoving box. The mass of each gas cell is $8.5\times10^4\ \Msun$, and the typical size of gas cells is around $100\ {\rm pc}$ in star-forming regions \citep{pillepich_first_2019}. 

The haloes in TNG50 are identified with the Friends-of-Friends algorithm, and the subhaloes are identified as gravitationally bound systems with the \textsc{subfind} algorithm \citep{springel_populating_2001}. We use the terms `galaxy' to refer to TNG50 `subhalo' hereafter. Once galaxies are identified, TNG50 applies the \textsc{sublink} algorithm \citep{rodriguez-gomez_merger_2015} to construct merger trees based on the identified galaxies. 

\subsection{Modeling cluster formation and evolution}
\label{sec:modeling_cluster_formation_and_evolution}

We model GC formation and evolution via three steps:
\begin{enumerate}
	\item \textit{Cluster formation:} calculate the total mass and metallicity of GCs based on the assembly history of the host galaxy.
	\item \textit{Cluster sampling:} compute the initial mass of each individual cluster and assign it to a collisionless particle.
	\item \textit{Cluster evolution:} evaluate the mass loss of clusters due to tidal disruption and stellar evolution. 
\end{enumerate}
Below we describe these three steps in detail.

\subsubsection{Cluster formation}
\label{sec:cluster_formation}

The cluster formation algorithm is similar to the previous versions of the model \citep{choksi_formation_2018,choksi_formation_2019,choksi_origins_2019}. 
First, we trigger a GC formation event when the galaxy mass grows suddenly. To quantify the rate of mass growth, we introduce the specific mass accretion rate, $R_{\rm m}$, as the fractional change of galaxy mass between two adjacent snapshots:
\begin{equation}
	R_{\rm m} = \frac{M_{\rm now} - M_{\rm prog}}{M_{\rm prog}}\cdot\frac{1}{t_{\rm now}-t_{\rm prog}}
	\label{eq:mass_accretion_rate}
\end{equation}
where $t_{\rm now}$ and $t_{\rm prog}$ stand for the cosmic times of the current snapshot and the progenitor snapshot, respectively. Similarly, the masses of the current galaxy and the progenitor galaxy are represented by $M_{\rm now}$ and $M_{\rm prog}$. If the current galaxy has more than one progenitor galaxies, we define $M_{\rm prog}$ as the main progenitor galaxy mass. GC populations form when $R_{\rm m}$ exceeds a threshold value, $p_3$, which is an adjustable parameter. It is worth noting that the \textsc{subfind} algorithm does not work robustly during mergers. In rare cases, the mass of the incoming galaxy may rise dramatically when it approaches the main galaxy, as a result of miss-identification. To fix this problem, we skip the snapshots when incoming galaxies suddenly gain mass during mergers. Once a GC formation event is triggered, we calculate the total mass of a newly formed GC population using the linear cluster mass--gas mass relation \citep{kravtsov_formation_2005}:
\begin{equation}
	M_{\rm tot} = 1.8\times 10^{-4} \, p_2 \, M_{\rm g}
	\label{eq:total_cluster_mass}
\end{equation}
where $M_{\rm g}$ is the cold gas mass of the host, and $p_2$ is another adjustable parameter\footnote{For consistency with previous work, we keep the notations of $p_2$ and $p_3$ as in \citet{li_modeling_2014}, although we introduce them in opposite order.}. The cold gas mass is approximated by the gas mass--stellar mass relation of \citet{choksi_formation_2018}:
\begin{equation}
	\eta(M_*, z)=\frac{M_{\rm g}}{M_*}=0.35\times 3^{2.7}\left(\frac{M_*}{10^9\Msun}\right)^{-n_M(M_*)}\left(\frac{1+z}{3}\right)^{n_z(z)}
	\label{eq:gas_mass_stellar_mass}
\end{equation}
based on the observations of \citet{lilly_gas_2013,genzel_combined_2015,tacconi_phibss_2018,wang_3_2022}. The mass-dependent power-law indices are
\begin{equation}
	n_M(M_*) = 	
	\left\{
		\begin{array}{lr}
		0.33, & {\rm for\;} M_* > 10^9 \Msun, \\
		0.19, & {\rm for\;} M_* < 10^9 \Msun,
		\end{array}
	\right.
\end{equation}
and the redshift dependency is characterized by
\begin{equation}
	n_z(z) = 
	\left\{
		\begin{array}{lr}
		1.4, & {\rm for\;} z > 2, \\
		2.7, & {\rm for\;} z < 2.
		\end{array}
	\right.
\end{equation}
Following \citet{choksi_formation_2018}, the metallicity of the newly formed cluster population is directly drawn from the metallicity of the interstellar medium of the host galaxy, which is also treated as a double power-law function of stellar mass and redshift:
\begin{equation}
	\feh=\log_{10}\left[\left(\frac{M_*}{10^{10.5}\Msun}\right)^{0.35}(1+z)^{-0.9}\right].
	\label{eq:metalicity_stellar_mass}
\end{equation}
We employ a $0.35$ slope for the stellar mass dependency as suggested by \citet{ma_origin_2016}. The $0.9$ slope of the redshift dependency accounts for the $0.6$~dex drop of $\feh$ from $z=0$ to $\sim4$ \citep{mannucci_lsd_2009}. Eqs.~(\ref{eq:gas_mass_stellar_mass}) and (\ref{eq:metalicity_stellar_mass}) both depend on the stellar mass of the host galaxy. To calculate the stellar mass from galaxy mass, we use a modified stellar mass--halo mass (SMHM) relation proposed by \citet{behroozi_average_2013}. The modified relation extends the original SMHM relation to $z>8$ and adds additional scatter to the original relation; see \citet{choksi_formation_2018} for detailed discussion. The galaxy mass is taken directly from the TNG50 catalogues.

Note that we apply analytic relations to calculate the gas mass, stellar mass, and metallicity of the host galaxy, although these values can be taken directly from a hydrodynamic simulation such as TNG50. However, different simulations employ different sub-grid prescriptions to model multiple physical processes. These sub-grid models, which are not the focus of our work, can significantly influence the formation and evolution of star clusters. By keeping minimal use of cosmological simulations, our model is less sensitive to such sub-grid models and can better reveal the link between GC systems and the assembly history of the host galaxy.

\subsubsection{Cluster sampling}
\label{sec:cluster_sampling}

To determine the number of newly formed clusters in a GC formation event, we stochastically sample clusters from a \citet{schechter_analytic_1976} initial cluster mass function (ICMF) with the `optimal sampling' method \citep{schulz_mass_2015}. Following \citet{choksi_formation_2019}, the truncation mass of ICMF is set to $\Mc=10^7\Msun$. 

Next, we link the newly formed GCs to collisionless particles in the simulation by the following assignment technique. Considering $N$ GCs that are formed within a galaxy at cosmic time $t_{\rm now}$, we randomly assign them to $N$ young stellar particles that belong to the galaxy. We set two constraints on candidate stellar particles to ensure they can correctly represent GCs. 

First, we select only inner particles as GCs since observations of young star clusters \citep[e.g.,][]{adamo_probing_2015,adamo_star_2020,randriamanakoto_young_2019} show that massive clusters preferentially form in the inner regions of host galaxies. As both observations \citep{van_der_wel_3d-hstcandels_2014,shibuya_morphologies_2015} and simulations \citep{pillepich_first_2019} suggest that the effective radii of stars in galaxies are around $1.5\ {\rm kpc}$ when GCs are actively forming ($z=2-5$), we include only stellar particles within twice the effective radius: $3\ {\rm kpc}$. This setup is also supported by observations of nearby dwarf galaxies that show the effective radii of their GC systems are only marginally larger than the effective radii of field stars \citep{carlsten_elves_2022}. Second, we select only stellar particles formed within a narrow time interval $\Delta t$ prior to $t_{\rm now}$. This interval characterizes a typical time for cluster formation and is set to $\Delta t = 10\ {\rm Myr}$.

For most cases, we have more than $N$ candidate particles that meet the two constraints, and we randomly select $N$ of them to represent GCs. However, in rare cases when there is an insufficient number of stellar particles formed within $\Delta t$, we adopt all of them and select the next most recently formed stellar particles until we have $N$ clusters. To prevent selecting stellar particles that are too old to be related to the GC formation event, we only adopt stellar particles younger than half of the time interval between adjacent outputs. In a very small number of cases, we still do not find enough stellar particles satisfying this criterion. In this case, we assign the remaining required number of GCs to dark matter particles located closest to the center of the dark matter halo, where the star-forming region is located. Only about $0.3\%$ of surviving GCs are represented by dark matter particles.

We do not use gas particles/cells (or `particles' in brief) for two reasons: 1) gas particles experience pressure forces and thus cannot correctly probe the kinematics of collisionless GCs; 2) in \textsc{arepo} simulations gas particles sometimes merge with other gas particles, making it difficult to trace them throughout cosmic time. 

\subsubsection{Cluster evolution}
\label{sec:cluster_evolution}

After the formation of GCs, we evaluate their mass loss due to tidal disruption and stellar evolution. The tidal disruption rate of a cluster with mass $M$ can be expressed as
\begin{equation}
	\frac{dM(t)}{dt}=-\frac{M(t)}{\ttid(M,t)}
	\label{eq:mass_loss_rate}
\end{equation}
where $\ttid$ is the tidal disruption timescale. As suggested by \citet{gieles_lifetimes_2008}, $\ttid$ depends significantly on the local tidal field parametrized by the orbital angular frequency, $\Omega_{\rm tid}$. In this work, we follow \citet{li_star_2019} to calculate $t_{\rm tid}$ as
\begin{equation}
	t_{\rm tid}(M,t) = 10\ {\rm Gyr} \left[\frac{M(t)}{2\times10^5\Msun}\right]^{2/3}\left[\frac{\Omega_{\rm tid}(t)}{100\ {\rm Gyr^{-1}}}\right]^{-1}.
	\label{eq:t_tid}
\end{equation}
The frequency $\Omega_{\rm tid}$ can be approximated by
\begin{equation}
	\Omega_{\rm tid}^2\simeq\Omega_{\lambda}^2\equiv\frac{\lambda_{\rm m}}{3} = \frac{\max|\lambda_i|}{3}.
	\label{eq:omega_tid}
\end{equation}
Variables $\lambda_i$ are the eigenvalues of the tidal tensor ${\bf T}(\mathbfit{x}_0,t)$, which is defined as
\begin{equation}
	T_{ij}(\mathbfit{x}_0,t) \equiv -\left.\frac{\partial^2\Phi(\mathbfit{x},t)}{\partial x_i\partial x_j}\right|_{\mathbfit{x}=\mathbfit{x}_0}
	\label{eq:tidal_tensor}
\end{equation}
where $i$ and $j$ are the orthogonal directions in the Cartesian coordinate system, and $\mathbfit{x}_0$ stands for the location of the cluster.

To numerically calculate the tidal tensor in the TNG50 simulation, we first place a $3\times 3\times 3$ cubic grid centered on the cluster. The side length of the grid is $2d$. The potential at each grid point is linearly interpolated from $8$ nearby particles with known potentials stored in the simulation snapshot. Finally, we approximate the diagonal terms of the tidal tensor via
\begin{equation}
    T_{ii}=-\frac{1}{d^2}[\Phi(\mathbfit{x}_0+\hat{\mathbfit{e}}_id)
    +\Phi(\mathbfit{x}_0-\hat{\mathbfit{e}}_id) 
    -2\Phi(\mathbfit{x}_0)],
\end{equation}
where $\hat{\mathbfit{e}}_i$ is the unit vector along the $i$ direction. Similarly, the non-diagonal terms are given by
\begin{multline}
    T_{ij}=-\frac{1}{4d^2}[\Phi(\mathbfit{x}_0+\hat{\mathbfit{e}}_id+\hat{\mathbfit{e}}_jd)
    +\Phi(\mathbfit{x}_0-\hat{\mathbfit{e}}_id-\hat{\mathbfit{e}}_jd) \\
    -\Phi(\mathbfit{x}_0+\hat{\mathbfit{e}}_id-\hat{\mathbfit{e}}_jd)
    -\Phi(\mathbfit{x}_0-\hat{\mathbfit{e}}_id+\hat{\mathbfit{e}}_jd)].
\end{multline}
Knowing the 9 terms of the tidal tensor, we can compute the three eigenvalues numerically. Plugging the eigenvalues into Eq.~(\ref{eq:omega_tid}), we get an estimate for $\Omega_{\lambda}$, denoted as $\tilde{\Omega}_{\lambda}$. It is important to choose a proper $d$ to calculate the tidal tensor accurately. A too-large $d$ tends to underestimate the tidal field in the central dense region of a galaxy, whereas a too small $d$ tends to overestimate the tidal field in the outer region where the density is lower. By performing a detailed test described in Appendix~\ref{sec:accuracy_of_approximating_tidal_disruption}, we suggest that $d=0.3\ {\rm kpc}$ can best approximate the tidal tensor for MW mass galaxies in TNG50.

An alternative approach to approximate $\Omega_{\rm tid}$ is to link the tidal tensor with the average mass density $\rho$ via Poisson's equation:
\begin{equation}
	4\pi G\rho(\mathbfit{x}_0,t) = \left.\nabla^2\Phi(\mathbfit{x},t)\right|_{\mathbfit{x}=\mathbfit{x}_0} = -{\rm tr}\left[{\bf T}(\mathbfit{x}_0,t)\right] = -\sum_i \lambda_i.
\end{equation}
Therefore, another approximation for $\Omega_{\rm tid}$ is given by
\begin{equation}
	\Omega_{\rm tid}^2\simeq\Omega_\rho^2\equiv\frac{4\pi G\rho}{3}.
\end{equation}
We introduce here a factor of $3$ such that $\Omega_\rho=\Omega_\lambda$ if $\lambda_{\rm m}=\sum_i\lambda_i=4\pi G\rho$, which happens for an isothermal density profile (see the analytical derivation in Appendix~\ref{sec:accuracy_of_approximating_tidal_disruption}). Numerically, the mass density is estimated by using a standard SPH kernel over all particle species. We denote the orbital angular frequencies given by this approach as $\tilde{\Omega}_{\rho}$.

We must approximate either the tidal tensor or the mass density on a spatial scale comparable to the tidal radius of GCs, i.e., at $20-50\ {\rm pc}$. However, this scale is beyond the spatial resolution of most cosmological simulations, including TNG50. To take into account systematic deviations between the actual orbital angular frequencies and the derived values, we introduce a new adjustable model parameter $\kappa$ as a correction:
\begin{equation}
	\Omega_{\lambda/\rho} = \kappa_{\lambda/\rho}\cdot\tilde{\Omega}_{\lambda/\rho}.
\end{equation}
Another important reason for introducing $\kappa$ is insufficient time resolution of simulations (there are only $20$ `full' snapshots in TNG50), which does not allow us to follow the tidal disruption in the initial phase after GC formation or during violent interactions. As suggested by high-resolution cluster formation simulations \citep[e.g.,][]{li_star_2019,li_formation_2022,meng_tidal_2022}, tidal disruption rate peaks at these rare phases. Calculating the tidal field only from the simulation snapshots usually ignores these phases and underestimates the disruption. Therefore, we need $\kappa>1$ to balance this underestimate. 

We find that $\Omega_{\rm tid}=\Omega_{\lambda}$ and $\Omega_{\rm tid}=\Omega_{\rho}$ produce GC catalogues with similar statistics in most aspects since the two estimates give similar values for most GCs, see Appendix~\ref{sec:accuracy_of_approximating_tidal_disruption}. For simplicity, we only display results from the $\Omega_{\rm tid}=\Omega_{\lambda}$ case throughout the rest of the work, unless specified otherwise.

Plugging $\Omega_{\rm tid}=\Omega_{\lambda/\rho}$ into Eq.~(\ref{eq:mass_loss_rate}) and (\ref{eq:t_tid}), we get the present-day mass of a GC due to tidal disruption as $M'(t)$. Assuming the time scale of stellar evolution is much shorter than $t_{\rm tid}$, the final mass of the GC is given by
\begin{equation}
	M(t) = M'(t)\left[1-\int_0^t\nu_{\rm se}(t')\ dt'\right],
	\label{eq:stellar_evolution}
\end{equation}
where $\nu_{\rm se}$ is the mass loss rate due to stellar evolution given by \citet{prieto_dynamical_2008}.

\section{Model Calibration}
\label{sec:model_calibration}

There are three adjustable parameters in our model: $p_2$ (Sec.~\ref{sec:cluster_formation}), $p_3$ (Sec.~\ref{sec:cluster_formation}), and $\kappa$ (Sec.~\ref{sec:cluster_evolution}). To find best values for the three parameters, we calibrate the model with observations. We run the model multiple times on $N_{\rm h}$ typical TNG50 galaxies with different $(p_2, p_3, \kappa)$ configurations to find the best one that minimizes a merit function. In Sections \ref{sec:observational_data} and \ref{sec:merit_function} we introduce the observational data and merit function, respectively. Next, we show the best parameter configurations for the two cases of tidal disruption: $\Omega_{\rm tid}=\Omega_{\lambda}$ and $\Omega_{\rm tid}=\Omega_{\rho}$.

\subsection{Observational data}
\label{sec:observational_data}

The observational data for extragalactic GC systems are the same as the data used in \citet{choksi_formation_2018}. They included samples from the Virgo Cluster Survey \citep[VCS,][]{peng_acs_2006}, 7 brightest cluster galaxies \citep[BCGs,][]{harris_globular_2014}, and M31 \citep{huxor_outer_2014}. To calibrate the kinematic properties of model GCs, we use the observations of the Galactic GC system, which has been extensively studied \citep[e.g.,][]{sollima_global_2017,sollima_eye_2019,baumgardt_accurate_2021,vasiliev_gaia_2021}. We use the Galactic GC catalogue\footnote{\url{https://people.smp.uq.edu.au/HolgerBaumgardt/globular/}} presented by \citet{hilker_galactic_2019}. This catalogue utilizes Hubble Space Telescope (HST) photometry and \textit{Gaia} EDR3 proper motions to provide phase space information for 162 Galactic GCs.

\subsection{Merit function}
\label{sec:merit_function}

We employ the following merit function to optimize model parameters:
\begin{equation}
	\mathcal{M} \equiv \frac{\chi_M^2}{N_{\rm h}} + \frac{1}{G_M} + \frac{2}{G_Z} + \left(\frac{\sigma_{Z}}{\overline{\sigma}_{Z}}\right)^2 + \frac{\chi^2_R}{N_{\rm h}} + \frac{\chi^2_\sigma}{N_{\rm MW}}.
\end{equation}
There are six terms in this function. The first three terms are identical to the merit function in \citet{choksi_formation_2018}. The first term is the reduced $\chi^2$ of the total mass of GC system at $z=0$, defined as
\begin{equation}
 	\frac{\chi_M^2}{N_{\rm h}} = \frac{1}{N_{\rm h}} \sum_h \frac{\left(\log_{10}M_{\rm GC}-\log_{10}M_{\rm obs}\right)^2}{0.35^2}
\end{equation}
where $M_{\rm GC}$ represents the total mass of GC system, and
\begin{equation}
	M_{\rm obs} = 3.4\times10^{-5} M_{\rm h},
\end{equation}
is the observed GC system mass--halo mass relation \citep{harris_dark_2015} with the scatter of $0.35\ {\rm dex}$. The sum is over $N_{\rm h}$ modeled halo systems.

The second and third terms represent the `goodness' of the present-day GC mass and metallicity distributions, respectively. Following \citet{li_modeling_2014}, we link observed galaxies to simulated galaxies with similar masses and compute the Kolmogorov-Smirnov (KS) test for each pair of linked galaxies. The terms $G_{\rm M}$ and $G_{\rm Z}$ represent the fraction of pairs with $p_{\rm KS}>0.01$, which can be taken as an acceptable match.

As suggested by \citet{choksi_formation_2018}, the model tends to underestimate the observed value of metallicity scatter $\sigma_{Z}=0.58\ {\rm dex}$. We introduce the fourth term as a penalty for such underestimation, where $\overline{\sigma}_{Z}$ is the average scatter of metallicity for modeled GC systems.

The last two terms are new in this work. The fifth term measures the deviation of modeled GC system size--galaxy mass relation from observations. \citet{hudson_correlation_2018} and \citet{forbes_how_2017} suggested that the effective radius of GC system can be described as a power-law function of the host galaxy mass. However, these two papers provided very different power-law indices of $0.88$ and $0.33$, based on partially-overlapping data sets. To reconcile this discrepancy, we combine the two data sets and perform a new power-law fit, which yields
\begin{equation}
	\log_{10}R_{\rm e} = 0.76+0.62\log_{10}\left(\frac{M_{\rm h}}{10^{12}\Msun}\right),
\end{equation}
where $R_{\rm e}$ is given in $\rm kpc$. The details of the fit are described later in Sec.~\ref{sec:radial_profiles}. Note that this relation is more commonly given in terms of the viral mass $M_{200}$ instead of the halo mass $M_{\rm h}$, although they are used interchangeably in many studies. We make a distinction here because $M_{200}$ and $M_{\rm h}$ are typically not the same in TNG50: $\log_{10}(M_{200}/M_{\rm h})$ varies from $0.05-0.25\ {\rm dex}$, with a mean value of $0.1\ {\rm dex}$. Therefore, we apply $\log_{10}(M_{200}/M_{\rm h})=0.1$ to connect the two masses throughout the work. Based on the $R_{\rm e}$--$M_{\rm h}$ relation, we introduce the fifth term as the reduced $\chi^2$ of modeled GC systems matching this relation:
\begin{equation}
	\frac{\chi_R^2}{N_{\rm h}} = \frac{1}{N_{\rm h}}\sum_{h} \frac{1}{0.22^2}\left[\log_{10}R_{\rm e}-0.62\left(\frac{M_{\rm h}}{10^{12}\Msun}\right)-0.76\right]^2,
\end{equation}
where $0.22\ {\rm dex}$ is the intrinsic scatter of the fit. 

Finally, the sixth term is the reduced $\chi^2$ for 3D velocity dispersion in MW mass galaxies, defined as galaxies with total masses between $10^{12}$ and $10^{12.2}\Msun$:
\begin{equation}
	\frac{\chi_\sigma^2}{N_{\rm MW}} = \frac{1}{N_{\rm MW}}\sum_{g} \frac{(\log_{10}\sigma_{\rm 3D}-\log_{10}\sigma_{\rm 3D,MW})^2}{0.2^2}.
\end{equation}
The 3D velocity dispersion takes into account all three dispersion components in the cylindrical coordinate system, $\sigma_{\rm 3D}=(\sigma_R^2+\sigma_{\phi}^2+\sigma_z^2)^{1/2}$. The axis of the cylindrical system is constructed along the net angular momentum vector of all stellar particles in the galaxy. We calculate $\sigma_{\rm 3D}$ as the total dispersion for all GCs in the galaxy. We also define $\sigma_{\rm 3D, MW}=200\ {\rm km\,s^{-1}}$ to represent the 3D velocity dispersion of the Galactic GC system. The intrinsic scatter of $\sigma_{\rm 3D}$ can be approximated by $0.2\ {\rm dex}$. 

\subsection{Parameter selection}
\label{sec:parameter_selection}

To calibrate our model with observations, we randomly pick $32$ central galaxies with total mass between $10^{11.5}-10^{12}\Msun$ and $32$ central galaxies with total mass between $10^{12}-10^{12.5}\Msun$ from the TNG50 simulation. There are $N_{\rm MW}=13$ galaxies that match our definition of MW mass galaxies, i.e., $\Mh = 10^{12}-10^{12.2}\Msun$. By minimizing the merit function $\mathcal{M}$ on the $64$ sample galaxies, we find a large region in the 3D parameter space can produce relatively good results with similar $\mathcal{M}$. In Appendix~\ref{sec:test_of_different_model_parameters}, we show that these different configurations only influence the final results sightly. Without loss of generality, we apply $(p_2, p_3, \kappa_\lambda)=(8, 0.5\ {\rm Gyr^{-1}}, 4)$ for the $\Omega_{\rm tid}=\Omega_{\lambda}$ case, and $(p_2, p_3, \kappa_\rho)=(8, 0.5\ {\rm Gyr^{-1}}, 5)$ for the $\Omega_{\rm tid}=\Omega_{\rho}$ case throughout the rest of the paper. To build some intuition about the value of $\kappa$, we provide an analytical estimate of the cluster lifetime for the $\Omega_{\rm tid}=\Omega_{\lambda}$ model. For a GC with initial mass $M=2\times 10^5\Msun$, a typical value of the tidal strength is $\lambdam=10^4\ {\rm Gyr^{-2}}$, corresponding to the distance of 3 kpc from the center of a MW-like galaxy at $z=0$, as shown in Fig.~\ref{fig:lambda_r_log}. This gives $\Omega_{\rm tid}=230\ {\rm Gyr^{-1}}$ for $\kappa_\lambda=4$. According to Eq.~(\ref{eq:t_tid}), the tidal disruption timescale of this cluster is 4.3 Gyr. Assuming $\Omega_{\rm tid}$ to be a constant and using Eq.~(\ref{eq:stellar_evolution}), we find that such a cluster would lose all its mass after 6.5 Gyr.

\section{Spatial and Kinematic Distributions}
\label{sec:comparison_with_observations}

In this section we describe new results and predictions made possible by inclusion of the spatial information in our model.

\subsection{Radial profiles}
\label{sec:radial_profiles}

The observed surface number density profiles of GC systems can be well fitted by the de Vaucouleurs law \citep{rhode_globular_2004,hudson_correlation_2018}, which is the $n=4$ case of the S\'{e}rsic profile
\begin{equation}
	\Sigma(R)=\Sigma_{\rm e}\exp\left(-b_n\left[\left(\frac{R}{R_{\rm e}}\right)^{1/n}-1\right]\right),
	\label{eq:sersic}
\end{equation}
where $R_{\rm e}$ stands for the effective radius, and $\Sigma_{\rm e}$ for the surface density at $R_{\rm e}$. The factor $b_n$ can be approximated by $b_n = 1.9992 \, n - 0.3271$ \citep{corwin_photometry_1989} for $0.5<n<10$.

\begin{figure}
	\includegraphics[width=\linewidth]{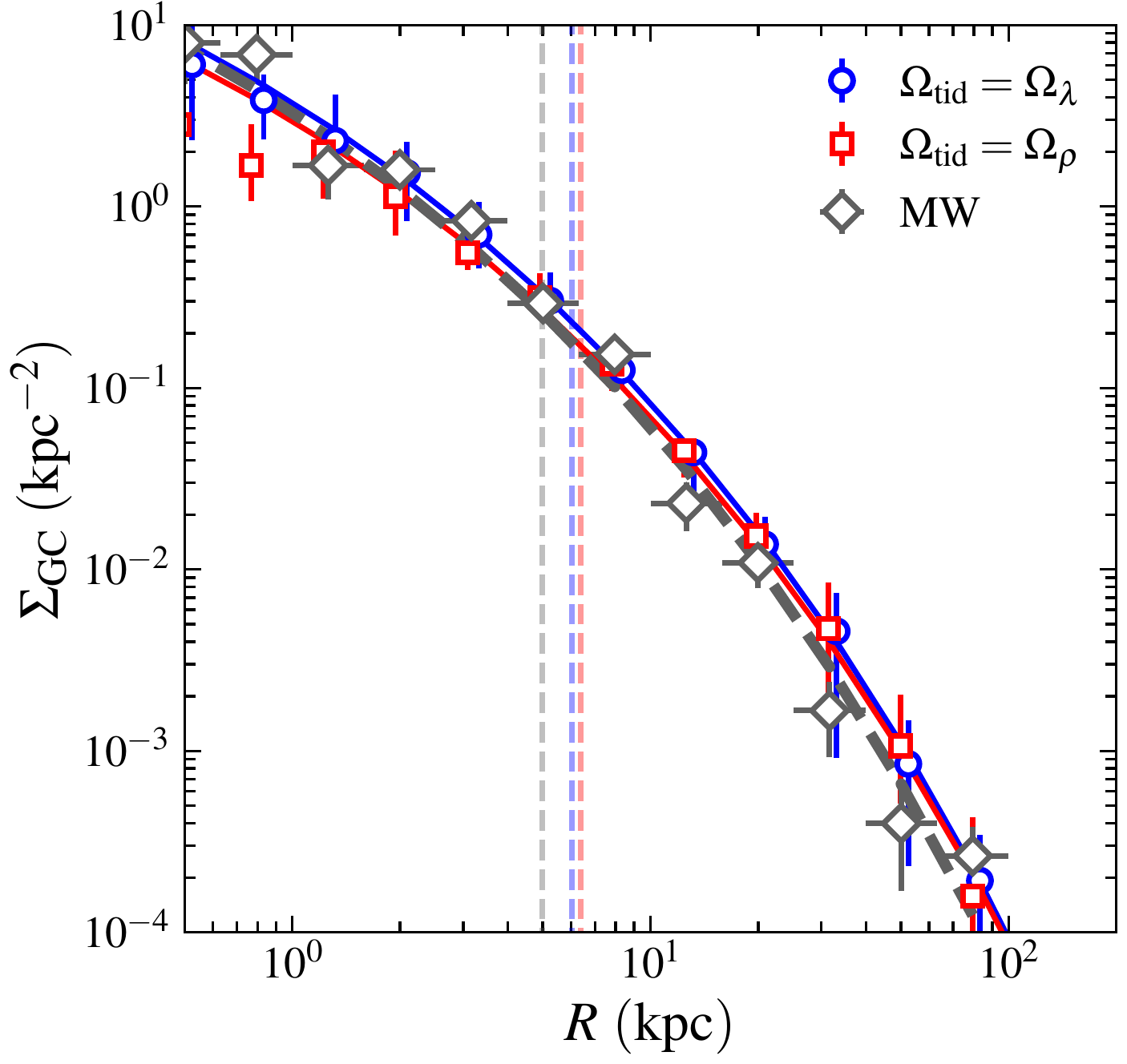}
	\vspace{-4mm}
    \caption{Surface number density profiles of model GC systems in MW mass galaxies. Models with the two implementations of tidal disruption are shown as blue circles ($\Omega_{\rm tid}=\Omega_{\lambda}$) and red squares ($\Omega_{\rm tid}=\Omega_{\rho}$), respectively. Symbols represent the median value of $13$ MW mass galaxies, while vertical errorbars correspond to the $16-84$th percentiles. For comparison, the surface number density profile of the MW GC system is shown as gray diamonds with errorbars: vertical errorbars show the Poisson error, horizontal errorbars indicate the bin width. Curves show the corresponding de Vaucouleurs fits, with vertical dashed lines marking the best-fit effective radii $R_{\rm e}$.}
    \label{fig:Sigma_r_log_mw}
\end{figure}

With the model GC systems projected onto the face-on planes of host galaxies, in Fig.~\ref{fig:Sigma_r_log_mw} we show the surface density profiles in the $13$ model MW mass galaxies at $z=0$. We define the face-on plane to be perpendicular to the net angular momentum vector for all stellar particles in the galaxy. The density profiles of both $\Omega_{\rm tid}=\Omega_{\lambda}$ and $\Omega_{\rm tid}=\Omega_{\rho}$ cases almost perfectly match the observed relation within the $16-84\%$ confidence level over a wide range of radii. Both models can be well fitted by the de Vaucouleurs law from $1$ to $100\ {\rm kpc}$ with $R_{\rm e}\sim 6\ {\rm kpc}$, which is close to the effective radius of the MW GC system, $R_{\rm e,MW}=5\ {\rm kpc}$.

\begin{figure}
	\includegraphics[width=\linewidth]{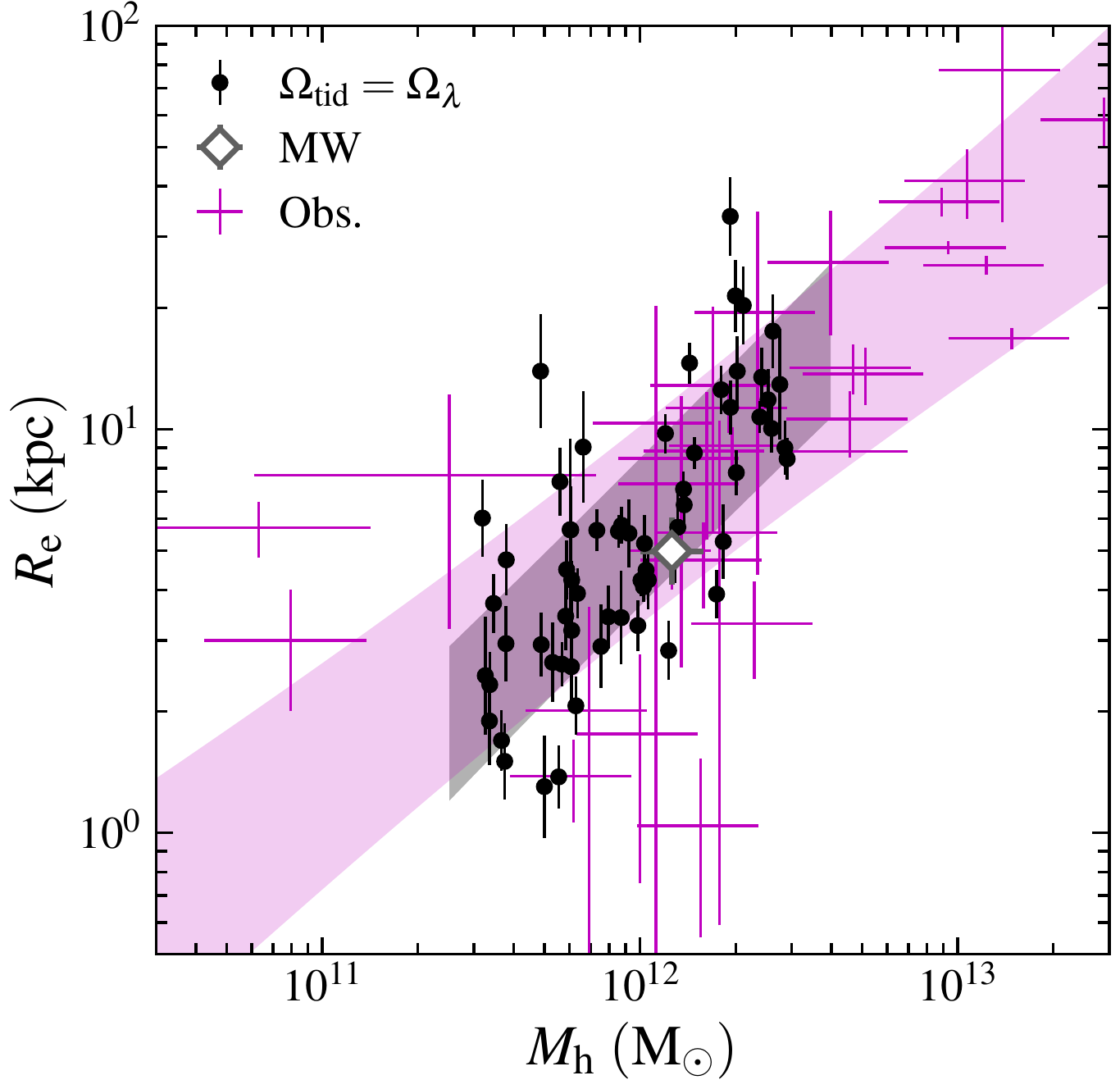}
	\vspace{-4mm}
    \caption{Best-fit de Vaucouleurs effective radii of modeled GC systems as a function of host halo mass, with fitting uncertainties of $R_{\rm e}$ shown as vertical errorbars. The observed GC systems from \citet{hudson_correlation_2018} and \citet{forbes_how_2017} are plotted as magenta crosses. The power-law fits from maximum likelihood for the model ($R_{\rm e}\propto M_{\rm h}^{0.79\pm 0.09}$) and observations ($R_{\rm e}\propto M_{\rm h}^{0.62\pm 0.13}$) are shown as the shaded regions, whose widths represent the 1-$\sigma$ confidence levels. MW is highlighted as the gray diamond: the horizontal errorbar shows the uncertainty of the total mass of MW: $10^{12}- 10^{12.2}\Msun$; while the vertical errorbar corresponds to the fitting error of $R_{\rm e, MW}$. Both errorbars are hard to observe as they are smaller than the marker size.}
    \label{fig:re_mh_log}
\end{figure}

To explore the spatial distribution of GCs in galaxies in a wider mass range, we plot the best-fit $R_{\rm e}$ as a function of the host halo mass for all sample galaxies in Fig.~\ref{fig:re_mh_log}. For clarity, we show the $R_{\rm e}$--$M_{\rm h}$ relation only for the $\Omega_{\rm tid}=\Omega_{\lambda}$ case, as all conclusions also stand for the $\Omega_{\rm tid}=\Omega_{\rho}$ case. Within the halo mass range of $10^{11.5}-10^{12.5}\Msun$, our model predicts the effective radii of GC systems to grow from $1$ to $10\ {\rm kpc}$, roughly following a power-law shape.

We perform a standard power-law fit using the maximum likelihood method, which takes into account the uncertainties in both $R_{\rm e}$ and $M_{\rm h}$ and the intrinsic scatter of the relation. The fit function reads
\begin{equation}
    \log_{10}R_{\rm e} = a + b\log_{10}M_{\rm h} + \epsilon,
    \label{eq:fit_function}
\end{equation}
where the intrinsic scatter is represented by a random variable $\epsilon$, which follows a Gaussian distribution ${\cal N}(0,\sigma_{\rm int})$. Correspondingly, the likelihood is given by
\begin{equation}
    {\cal L}\equiv\prod_i \frac{1}{\sigma_i\sqrt{2\pi}}\exp\left(-\frac{\delta_i^2}{2\sigma_i^2}\right),
    \label{eq:likelihood}
\end{equation}
where $\sigma_i^2=\sigma_{\log R,i}^2+b^2\sigma_{\log M,i}^2+\sigma_{\rm int}^2$, and $\delta_i=\log_{10}R_{{\rm e},i}-a-b\log_{10}M_{{\rm h},i}$ is the `vertical' deviation. We introduce $\sigma_{\log R,i}$ and $\sigma_{\log M,i}$ as the observed uncertainties of $\log_{10}R_{{\rm e},i}$ and $\log_{10}M_{{\rm h},i}$, respectively, with subscript $i$ corresponding to the $i$-th data point. Additionally, we apply bootstrap resampling $5000-10000$ times until all fitting parameters converge to estimate the standard deviations of $a$, $b$, and $\sigma_{\rm int}$, denoted by $\sigma_a$, $\sigma_b$, and $\sigma_\sigma$. By assuming the fitting parameters to be random variables following Gaussian distributions (e.g., the slope follows ${\cal N}(b,\sigma_b)$), the predicted $\log_{10}R_{\rm e}$ can also be described as a Gaussian distribution, whose mean value is given by Eq.~(\ref{eq:fit_function}), and uncertainty is given by $\sigma_{\log R}^2=\sigma_a^2+(\sigma_b\log_{10}M_{\rm h})^2+\sigma_{\rm int}^2$. For our model data, the uncertainty of $\log_{10} M_{\rm h}$ is set to zero since $M_{\rm h}$ is directly taken from TNG50. Maximizing the likelihood ${\cal L}$ yields
\begin{equation}
    \log_{10}R_{\rm e} = (0.75\pm0.03) + (0.79\pm0.09)\log_{10}\left(\frac{M_{\rm h}}{10^{12}\Msun}\right),
    \label{eq:simple_fit_model}
\end{equation}
where $R_{\rm e}$ is given in kpc. We also obtain an intrinsic scatter of $\sigma_{\rm int}=0.18\pm0.02$. 

By minimizing the `vertical' deviations $\delta_i$, the above fit assumes one variable to be dependent ($R_{\rm e}$) and the other independent ($M_{\rm h}$). Alternatively, one may perform an orthogonal fit to the two-dimensional distribution, treating both variables as independent. It describes situations where both variables are the result of many complex processes, such as those operating in galaxy formation. With the same fit function, Eq.~(\ref{eq:fit_function}), the orthogonal fit aims to minimize the perpendicular deviation of data points from the best-fit line. The likelihood function in this case is
\begin{equation}
    {\cal L}_{\rm ortho}\equiv\prod_i\frac{1}{\Sigma_i\sqrt{2\pi}}\exp\left(-\frac{\Delta_i^2}{2\Sigma_i^2}\right).
\end{equation}
The perpendicular deviation is $\Delta_i=(\log_{10}R_{{\rm e},i}-a)\cos\theta-M_{{\rm h},i}\sin\theta$, where $\theta=\tan^{-1}b$ is the inclination angle. Similarly, we write the perpendicular uncertainty $\Sigma_i^2=\sigma_{\log R,i}^2\cos^2\theta+\sigma_{\log M,i}^2\sin^2\theta+\sigma_{\rm int}^2\cos^2\theta$. Maximizing this likelihood ${\cal L}_{\rm ortho}$ for the modeled GCs relation yields
\begin{equation}
    \log_{10}R_{\rm e}^{\rm ortho} = (0.77\pm0.06) + (1.29\pm0.55)\log_{10}\left(\frac{M_{\rm h}}{10^{12}\Msun}\right)
\end{equation}
with an intrinsic scatter $\sigma_{\rm int}=0.26\pm0.16$. The slope of the orthogonal fit is significantly higher than that of the standard fit, as noted by various studies \citep[e.g.,][]{linnet_evaluation_1993}. The discrepancy is due to the two methods handling uncertainties differently. In general, there is a priori preference for either method: the standard fit can more clearly show how the dependent variable changes with independent variables, while the orthogonal fit is more appropriate to show the relation between two independent variables. Since we start with known $M_{\rm h}$ and model the prediction for $R_{\rm e}$, the standard fit is more appropriate in this work. We also provide the results from orthogonal fit for completeness.

We compare our model results with the observations of extragalactic GC systems presented in \citet{forbes_how_2017} and \citet{hudson_correlation_2018}. These samples include early type galaxies, massive cD galaxies, and three ultra diffuse galaxies (UDGs). Many galaxies overlap in the two samples. We combine the galaxies in both samples, exclude galaxies without any measure of halo mass, and update the mass estimate of DF 44. \citet{saifollahi_number_2021} suggested that DF44 is much less massive and has a more compact GC system than previously believed. It is worth noting that the masses of the three UDGs are obtained differently from other galaxies: the masses of UDGs are inferred from their GC mass/counts, while the masses of most other galaxies are derived from the stellar mass--halo mass relation. This may introduce systematic errors when fitting them with a single power-law relation.

We show the $R_{\rm e}$--$M_{\rm h}$ relation for this combined observational sample in Fig.~\ref{fig:re_mh_log} and note that the observational data roughly follow a power-law relation. By performing a fit on these observational data with the standard likelihood, we obtain
\begin{equation}
    \log_{10}R_{\rm e, obs} = (0.76\pm0.10) + (0.62\pm0.13)\log_{10}\left(\frac{M_{\rm h}}{10^{12}\Msun}\right)
    \label{eq:simple_fit_obs}
\end{equation}
with an intrinsic scatter $\sigma_{\rm int}=0.22\pm0.06$. The slope of $0.62$ lies between the values of $0.88$ and $0.33$ quoted by \citet{hudson_correlation_2018} and \citet{forbes_how_2017}, respectively. Note that Eq.~(\ref{eq:simple_fit_obs}) gives $R_{\rm e}=6.6$ kpc for a MW mass ($10^{12.1}\Msun$) system. In comparison, the MW GC system has $R_{\rm e}=5$ kpc, which is by $0.12$ dex more compact than the average MW mass system. It is not surprising that the model also tends to overestimate the size of MW GC system by similar amount since the model is calibrated with the $R_{\rm e}$--$\Mh$ relation by Eq.~(\ref{eq:simple_fit_obs}). Moreover, this overestimation can be adjusted by applying lower strengths of tidal disruption (i.e., lower $\kappa$, as discussed in Appendix.~\ref{sec:test_of_different_model_parameters}) or different schemes of tidal disruption (e.g., constant disruption rate, as discussed in Sec.~\ref{sec:comparison_with_constant_disruption_rate}).

Alternatively, the orthogonal fit gives
\begin{equation}
    \log_{10}R_{\rm e,obs}^{\rm ortho} = (0.66\pm0.20) + (0.83\pm0.31)\log_{10}\left(\frac{M_{\rm h}}{10^{12}\Msun}\right)
\end{equation}
with an intrinsic scatter $\sigma_{\rm int}=0.26\pm0.17$. Similarly to the fits for modeled systems, the slope of the orthogonal fit is significantly higher than that of the standard fit. 

Considering the uncertainties of fitting parameters and large intrinsic scatter, the model results and observations are consistent with each other, although the two slopes are formally different. We will investigate the power-law slope of the $R_{\rm e}$--$M_{\rm h}$ relation further in Sec.~\ref{sec:comparison_with_constant_disruption_rate}, where we focus on how tidal disruption alters the sizes of GC systems. We find that a model with location-sensitive disruption (such as in this work) tends to have steeper $R_{\rm e}$--$M_{\rm h}$ relation compared with location-independent models.

\subsection{Kinematics}
\label{sec:kinematics}

Since now we have observational measurements of 3D velocities for most Galactic GCs, we can compare them with the kinematics of the modeled GC systems for MW mass galaxies. We project the velocity of each GC onto the cylindrical coordinate system centered on the host galaxy and calculate the three perpendicular components: the radial component, $v_R$; the azimuthal component, $v_\phi$; and the axial component, $v_z$. The axis of the coordinate system is aligned with the net angular momentum vector of all stellar particles in the galaxy. For each of these components, we define velocity dispersions as the standard deviations after subtracting the mean.

\begin{figure}
	\includegraphics[width=\linewidth]{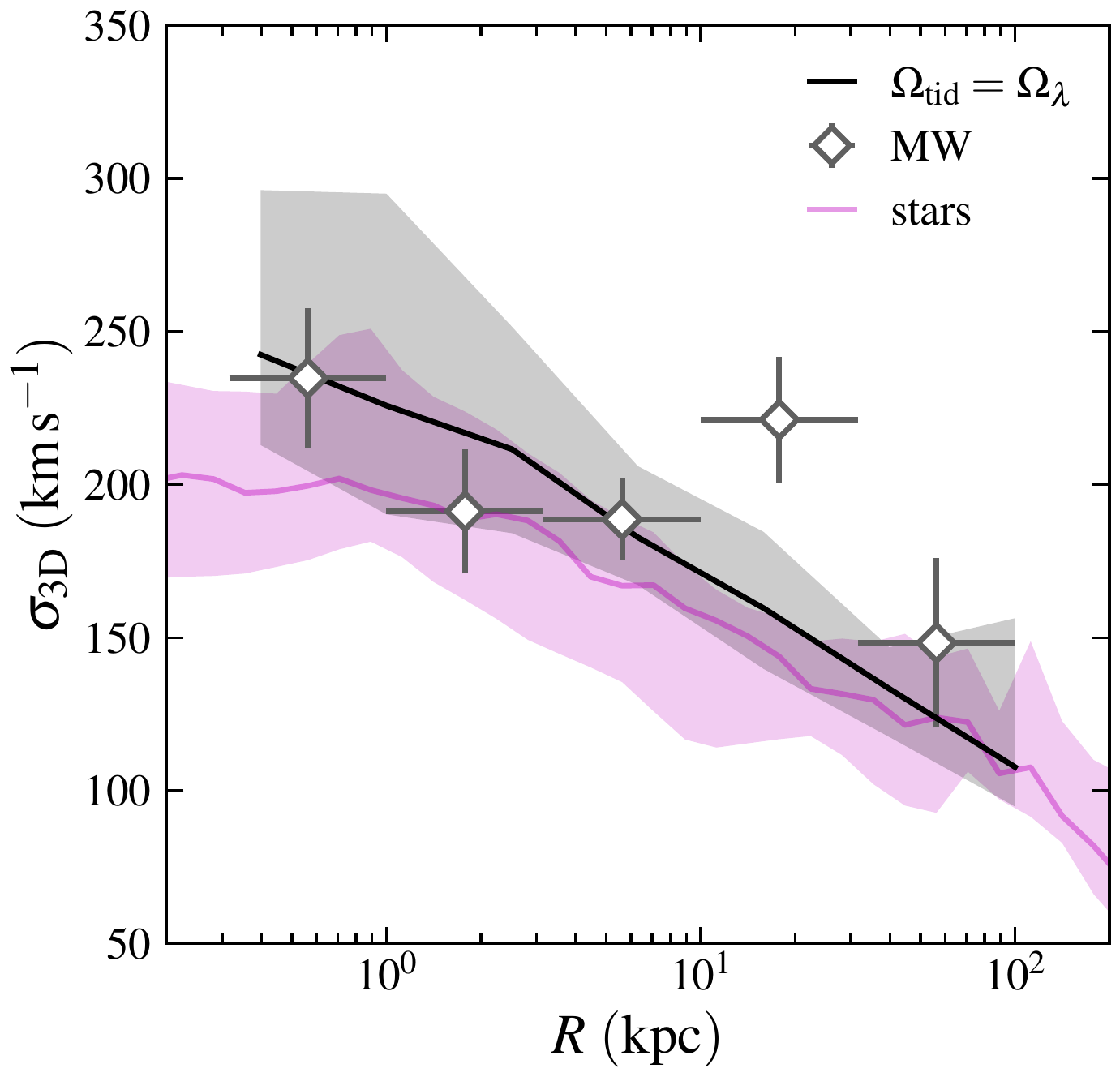}
	\vspace{-5mm}
    \caption{Radial profile of 3D velocity dispersion for model GC systems in MW mass galaxies (black curve). The dispersion profile for all stellar particles in TNG50 galaxies is shown by the magenta curve. Shaded regions correspond to the $16-84$th percentiles of 13 MW mass galaxies. The observed dispersion profile of the MW GC system is shown as gray diamonds with errorbars: vertical errorbars represent the $16-84\%$ confidence levels from bootstrap resampling, and horizontal errorbars correspond to the bin width. We repeat bootstrap resampling 1000 times until the estimated confidence levels converge.}
    \label{fig:dispersion_r}
\end{figure}

Fig.~\ref{fig:dispersion_r} shows the radial profiles of the 3D velocity dispersion, $\sigma_{\rm 3D}=(\sigma_R^2+\sigma_{\phi}^2+\sigma_z^2)^{1/2}$. The dispersion is a decreasing function of radius: between 1 and $100\ {\rm kpc}$, $\sigma_{\rm 3D}$ drops from 230 to $100\kms$, with a scatter of $20-50\kms$. The model $\sigma_{\rm 3D}$ profile is in agreement with the observations of the Galactic GCs except at $R\simeq10-30$~kpc, where the observed dispersion jumps dramatically to $220\kms$. This bump is mainly created by the GCs associated with the Sagittarius stream \citep{vasiliev_proper_2019}. These GCs with similar radii have large velocities, which significantly inflates the velocity dispersion at $R\simeq30$~kpc. In fact, such random bumps are not rare in the modeled systems, but we do not observe any obvious bump in Fig.~\ref{fig:dispersion_r} since the uncertainties of the $\sigma_{\rm 3D}$ profile are represented by $16-84$th percentiles. For the 13 MW-mass galaxies, we can observe a clear bump only if there are more than 2 galaxies presenting bumps at the same radius, which is rather rear. It is worth noting that the good agreement with observations is not a trivial outcome of optimizing the merit function (Sec.~\ref{sec:merit_function}). We emphasize that the model result accurately reproduces the observed radial distribution of $\sigma_{\rm 3D}$, while the merit function only takes into account the total $\sigma_{\rm 3D}$ of all GCs in a galaxy.

We also compare the GC dispersion profile with the dispersion profile of all stellar particles. The velocity dispersion of GCs is about $10\%$ higher than that of the stellar component at the same radius, suggesting that GCs are more supported by random motion. However, this difference is smaller than the scatters and may be difficult to detect in observations.

\begin{figure}
	\includegraphics[width=\linewidth]{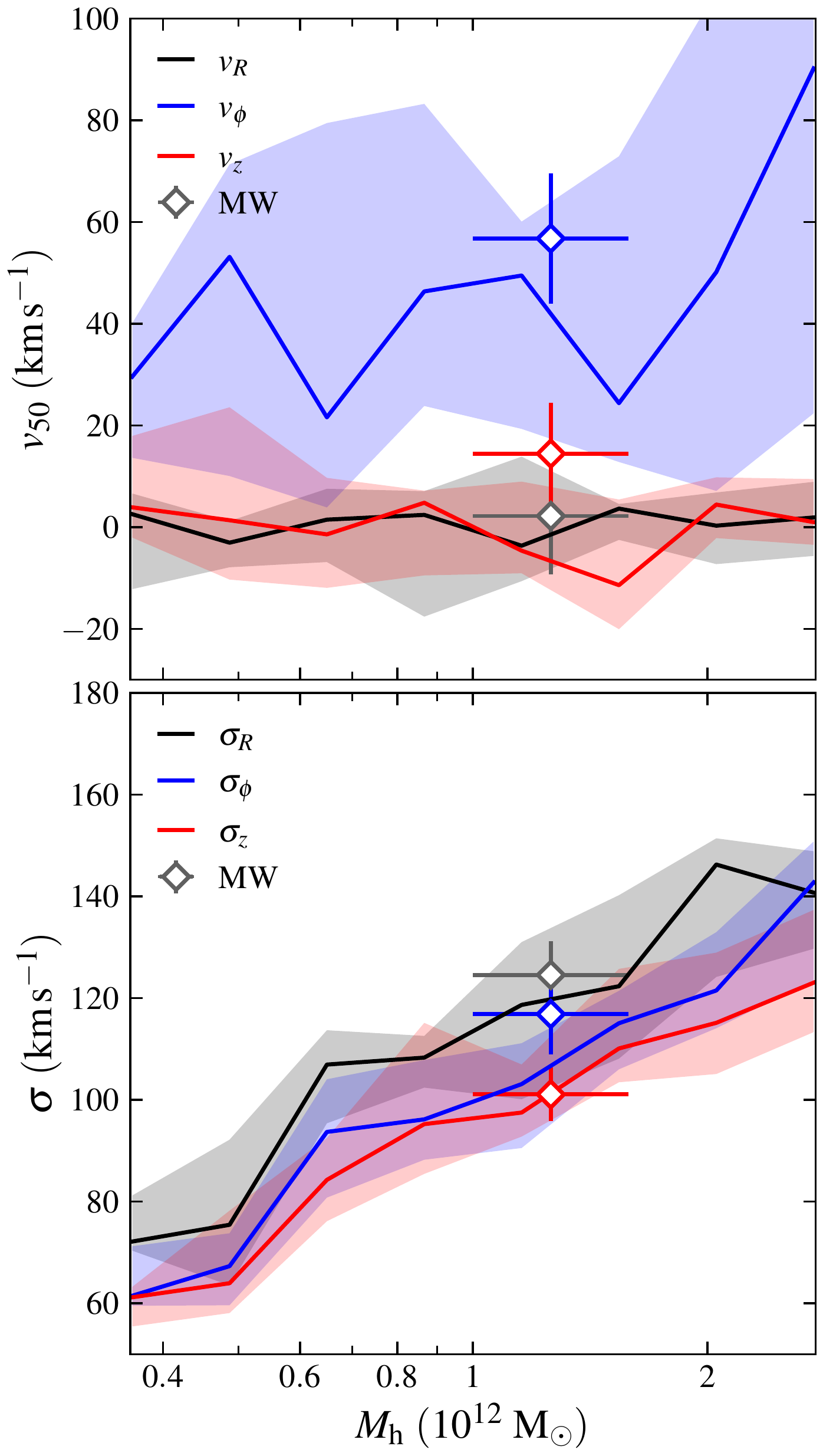}
	\vspace{-5mm}
    \caption{Median systemic velocity components (\textit{top panel}) and velocity dispersions (\textit{bottom panel}) as functions of host halo mass for model GC systems. Curves with shaded regions represent the $16-50-84$th percentiles of each component (note that the curves in the \textit{top panel} are medians of median values). The median systemic velocities and velocity dispersions of the Galactic GC system are overplotted as diamonds with errorbars: vertical errorbars represent uncertainties via bootstrap resampling, and horizontal errorbars represent the uncertainty of the MW mass.}
    \label{fig:v_sigma_mh_log}
\end{figure}

In addition to the dispersion profiles of GCs in MW mass galaxies, we investigate the kinematics of GC systems in galaxies of other mass. The top panel of Fig.~\ref{fig:v_sigma_mh_log} shows the median velocity components for all modeled galaxies as a function of halo mass. For all components, the median velocities are insensitive to the halo mass between $M_{\rm h}=10^{11.5}-10^{12.5}\Msun$. The median $v_R$ and $v_z$ of model GC systems are consistent with zero, with a rather small scatter of $\sim 10\kms$. On the other hand, the median azimuthal (rotational) velocity is $v_{\phi,50}=(40\pm30)\kms$, which is systematically non-zero with larger scatter. This reveals that the GC system is also rotating alongside the stellar component of the host galaxy. For quantitative comparison with the MW, we present in Table~\ref{tab:properties} the $16-50-84\%$ values of the three median velocity components for the 13 modeled MW mass galaxies as well as the MW GC systems. In addition, we list also the percentiles of MW properties in the modeled MW mass galaxies. These percentiles indicate how much the model can represent the MW properties: a percentile $>50$ ($<50$) means that this property of the MW is greater (smaller) than the median value of the 13 MW mass galaxies. For statistical significance, only percentiles $>84$ ($<16$) can be interpreted as the model systematically underestimating (overestimating) the corresponding properties of the MW. The model predictions of median $v_R$ and $v_\phi$ are consistent with the observations as these two properties of the MW overlap the $15-90\%$ and $53-93\%$ (intersecting with the $16-84\%$ range) values of the modeled values, respectively. Although the median $v_{z,{\rm MW}}$ of $19_{-11}^{+7}\kms$ is systematically non-zero, which is inconsistent with the model prediction that median $v_z$ is around zero, we still find some model systems to have an even greater median $v_z$ as the percentile of $v_{z,{\rm MW}}$ in model systems is smaller than $100$. It is not surprising that the systemic velocities can deviate from the average values since different dynamical histories of galaxies can lead to different kinematics of GCs. For example, \citet{vasiliev_proper_2019} suggested that the Sagittarius stream clusters have high polar velocities, which can significantly alter the distribution of systemic velocities. 

\begin{table*}
 \caption{Summary of key properties of GC systems in MW mass galaxies: median systemic velocities $v_{50}$, velocity dispersions $\sigma$, median orbital actions $J_{50}$, effective radii $R_{\rm e}$, and pericenter/apocenter radii $r_{50}$. All values and lower/upper uncertainties represent the median and 16$-$84th percentiles, respectively. The first three rows show the model results for 13 MW mass galaxies with the $\Omega_{\rm tid}=\Omega_\lambda$ prescription. The middle three rows show the observed values for the MW GC system. The uncertainties for the MW are calculated via bootstrap resampling; the progenitors of MW GCs are classified by \citet{massari_origin_2019}. The bottom three rows show the percentiles of the MW properties within the sample of 13 modeled galaxies. For example, a percentile $>50$ ($<50$) means that the MW property is greater (smaller) than the median of modeled values, i.e., the model tends to underestimate (overestimate) this property for the MW. Percentiles close to 0 or 100 indicate discrepancies between the model sample and the observations.}
 \label{tab:properties}
 \renewcommand\arraystretch{1.2}
 \begin{tabular}{llcccccccccccc}
  \hline
   & & \multicolumn{3}{c}{$v_{50}\ (\kms)$} & 
   \multicolumn{3}{c}{$\sigma\ (\kms)$} & 
   \multicolumn{3}{c}{$J_{50}\ ({\rm kpc\,\kms})$} &
   \multicolumn{3}{c}{$r_{50}\ ({\rm kpc})$} \\
   & & $v_R$ & $v_\phi$ & $v_z$ & 
   $\sigma_R$ & $\sigma_\phi$ & $\sigma_z$ & 
   $J_R$ & $|J_\phi|$ & $J_z$ & 
   $R_{\rm e}$ & $r_{\rm peri}$ & $r_{\rm apo}$\\
  \hline
  Model & all & 
    $0_{-10}^{+5}$ & $43_{-31}^{+17}$ & $-5_{-10}^{+11}$ &
    $122_{-20}^{+12}$ & $104_{-14}^{+11}$ & $103_{-11}^{+10}$ &
    $263_{-29}^{+197}$ & $421_{-111}^{+354}$ & $219_{-54}^{+200}$ &
    $5.2_{-1.0}^{+3.6}$ & $2.7_{-1.2}^{+1.1}$ & $11.7_{-3.5}^{+6.7}$ \\
   & \textit{in-situ} & 
    $1_{-14}^{+7}$ & $53_{-38}^{+47}$ & $-1_{-13}^{+18}$ & 
    $105_{-35}^{+34}$ & $102_{-33}^{+28}$ & $94_{-33}^{+25}$ & 
    $101_{-27}^{+102}$ & $264_{-73}^{+36}$ & $111_{-47}^{+55}$ &
    $3.0_{-0.6}^{+1.2}$ & $1.5_{-0.3}^{+0.5}$ & $5.5_{-1.5}^{+2.3}$ \\
   & \textit{ex-situ} & 
    $0_{-12}^{+13}$ & $31_{-28}^{+45}$ & $1_{-13}^{+15}$ & 
    $115_{-30}^{+28}$ & $95_{-28}^{+28}$ & $94_{-28}^{+30}$ & 
    $613_{-184}^{+299}$ & $682_{-258}^{+547}$ & $447_{-152}^{+223}$ &
    $9.9_{-1.4}^{+8.3}$ & $3.8_{-1.3}^{+1.6}$ & $19.3_{-4.1}^{+17.5}$ \\
  \hline
  MW & all & 
    $2_{-12}^{+13}$ & $57_{-12}^{+15}$ & $14_{-7}^{+11}$ &
    $124_{-7}^{+7}$ & $116_{-8}^{+8}$ & $101_{-5}^{+5}$ &
    $176_{-32}^{+73}$ & $342_{-19}^{+52}$ & $179_{-18}^{+24}$ &
    $5.0_{-0.8}^{+1.1}$ & $1.6_{-0.1}^{+0.2}$ & $7.0_{-0.6}^{+0.8}$ \\
   & \textit{in-situ} & 
    $2_{-15}^{+15}$ & $116_{-11}^{+6}$ & $8_{-7}^{+10}$ & 
    $95_{-8}^{+8}$ & $110_{-10}^{+9}$ & $89_{-8}^{+7}$ & 
    $59_{-9}^{+19}$ & $333_{-127}^{+61}$ & $76_{-12}^{+13}$ &
    $2.3_{-0.2}^{+0.3}$ & $1.5_{-0.3}^{+0.1}$ & $4.1_{-0.5}^{+0.2}$ \\
   & \textit{ex-situ} & 
    $2_{-18}^{+13}$ & $24_{-14}^{+11}$ & $32_{-24}^{+16}$ & 
    $143_{-9}^{+9}$ & $114_{-14}^{+12}$ & $109_{-7}^{+7}$ & 
    $825_{-148}^{+273}$ & $369_{-42}^{+47}$ & $357_{-35}^{+40}$ &
    $8.6_{-1.1}^{+0.9}$ & $1.9_{-0.3}^{+0.2}$ & $18.7_{-3.7}^{+1.6}$ \\
  \hline
  Percentile & all & 
    15$-$90 & 53$-$93 & 84$-$99 &
    41$-$80 & 61$-$98 & 29$-$59 &
    0$-$38 & 23$-$47 & 15$-$41 &
    14$-$63 & 16$-$23 & 0$-$4 \\
   & \textit{in-situ} & 
    17$-$91 & 85$-$91 & 53$-$86 & 
    30$-$49 & 47$-$75 & 32$-$56 & 
    0$-$21 & 22$-$92 & 16$-$30 &
    10$-$21 & 15$-$70 & 7$-$27 \\
   & \textit{ex-situ} & 
    12$-$86 & 25$-$55 & 71$-$99 & 
    74$-$93 & 60$-$89 & 63$-$77 & 
    54$-$92 & 0$-$12 & 29$-$43 & 
    9$-$45 & 0$-$10 & 12$-$52 \\
  \hline
 \end{tabular}
\end{table*}

In the bottom panel of Fig.~\ref{fig:v_sigma_mh_log} we compare the velocity dispersion of each component for the modeled galaxies with the MW GCs. We find that the dispersion of all three components increases significantly with halo mass. From $M_{\rm h}=10^{11.5}$ to $10^{12.5}\Msun$, the three components rise from $\sim$70 to $\sim$130$\kms$. As presented in Table~\ref{tab:properties}, the model predictions are in good agreement with observations as the observed median $\sigma_R$, $\sigma_\phi$, and $\sigma_z$ overlap the $41-80\%$, $61-98\%$, and $29-59\%$ values of the modeled MW mass galaxies, respectively; all percentiles intersect with the $16-84\%$ confidence level.

\begin{figure}
	\includegraphics[width=\linewidth]{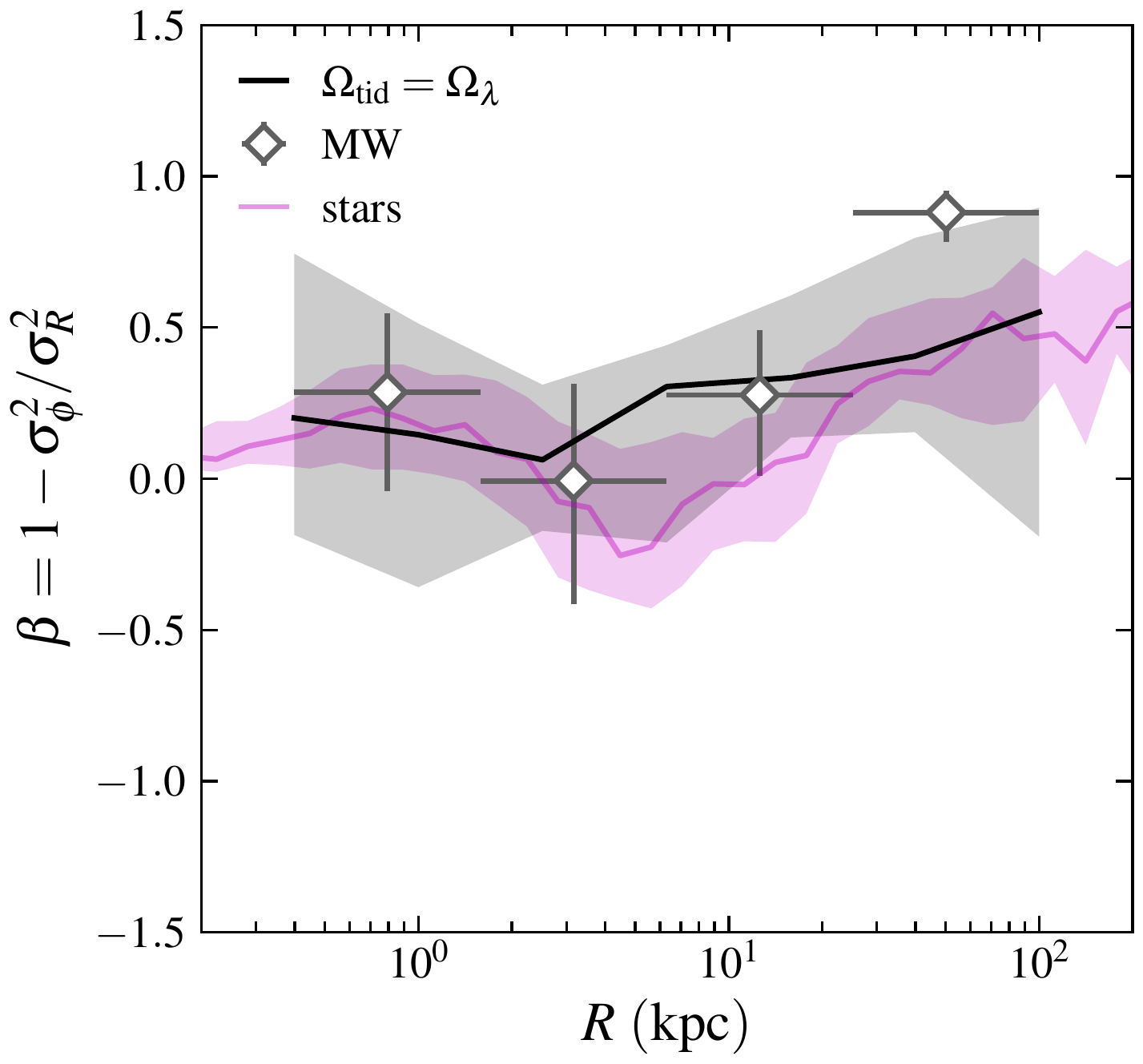}
 	\vspace{-4mm}
    \caption{Radial profiles of anisotropy parameter for modeled GC systems in MW mass galaxies. The anisotropy parameter profile of all stellar particles in TNG50 galaxies is shown by the magenta curve. Shaded regions represent the $16-84$th percentiles of the corresponding components. We also show the anisotropy parameter profiles for the MW GC system as diamonds with errorbars: vertical errorbars represent the $16-84\%$ confidence levels via bootstrap resampling, and the horizontal errorbars correspond to the bin width.}
     \label{fig:beta_r_log}
\end{figure}

We note that the radial dispersion $\sigma_R$ is generally greater than the tangential component $\sigma_\phi$, for galaxies with $M_{\rm h}=10^{11.5}-10^{12.5}\Msun$. Quantitatively, the anisotropy of radial and tangential motions can be characterized by the anisotropy parameter \citep{binney_radius-dependence_1980}. Here, we apply the definition of anisotropy parameter in cylindrical system by \citet{tonry_anisotropic_1983}, $\beta \equiv 1-\sigma_\phi^2/\sigma_R^2$. Negative (positive) values of $\beta$ correspond to a tangentially (radially) anisotropic velocity distribution. The case of $\beta=0$ corresponds to an isotropic velocity distribution. We plot the radial profile of $\beta$ for model GC systems in Fig.~\ref{fig:beta_r_log}. The model $\beta$ profile for all GCs rises gradually from $\beta \simeq 0$ to $0.5$ at $R = 1-100\ {\rm kpc}$. The model profile mostly matches the observed profile within the $16-84\%$ confidence level. Although model GCs are mainly represented by stellar particles, the entire stellar components in TNG50 galaxies tend to acquire lower $\beta$ values at $R = 2-20\ {\rm kpc}$, indicating that the GCs at this region are in general more radially biased than the field stars. Additionally, unlike the increasing $\beta$ of GCs, the stellar $\beta$ decreases with radius until $R \simeq 5\ {\rm kpc}$, where the anisotropy profile of field stars shows a dip, which may be related to past mergers of galaxies, as suggested by \citet{loebman_beta_2018}. 

\subsection{Orbital actions}

Recent studies \citep[][]{trujillo-gomez_kinematics_2021,wu_using_2021,callingham_chemo-dynamical_2022} focusing on the kinematics of GCs and halo stars have shown that orbital actions and integral of motions are useful probes of the dynamical histories of galaxies. These quantities are generally conserved during the slow evolution of the gravitational potential \citep{binney_galactic_2008}. To study the orbits of our GCs, we use the \textsc{agama} package \citep{vasiliev_agama_2019} to compute their pericenter and apocenter radii ($r_{\rm peri}$, $r_{\rm apo}$) and orbital actions ($J_R$, $J_\phi$, $J_z$). The actions of a closed orbit are defined as
\begin{equation}
    J_q=\frac{1}{2\pi}\oint\frac{p_q}{m}dq,
    \label{eq:action}
\end{equation}
where $q\in\{R,\phi,z\}$ corresponds to the radial, azimuthal, and vertical coordinates in a cylindrical system. The actions have the same dimension as the specific angular momentum. In fact, the azimuthal action $J_\phi$ is equivalent to the specific angular momentum along the $z$-axis, $L_z$. The sum $|J_\phi|+J_z$ equals the total specific angular momentum $L$.

Orbit calculation requires analytical modeling of the gravitational potential of TNG50 galaxies. We achieve it by employing the \textsc{agama} functionality, which approximates the potential with the multipole expansion method \citep[for details, see][]{vasiliev_agama_2019}. Since the potential of MW can be described by spheroids and disks \citep[e.g.,][]{mcmillan_mass_2017}, we model the present-day potentials of TNG50 galaxies with these two components. A largely spherical dark matter potential is modeled by spherical harmonic expansion, while the disky baryonic (star $+$ gas) potential is modeled by azimuthal harmonic expansion. The radial and vertical coordinates in the two expansion schemes are approximated by quintic splines. We find the multipole expansion approximation to be accurate, as it deviates from the simulation-provided potential by less than 2\%. Such a deviation is so small that we can ignore its influence on the subsequent calculation of orbital parameters. Next, we input the $z=0$ positions and velocities of GCs to \textsc{agama} and perform orbit integration to obtain the pericenter/apocenter radii and orbital actions. Note that these parameters are observable since we can apply the same procedure to MW GCs with the full 3-dimensional positions and velocities, assuming the \citet{mcmillan_mass_2017} model for the MW potential.

We list in Table~\ref{tab:properties} the pericenter and apocenter radii for the modeled systems as well as the MW GCs. The modeled values of the two radii are greater than the observed median values by $\sim0.2$~dex. This is because the model is calibrated with Eq.~(\ref{eq:simple_fit_obs}), where the MW GC system is more compact by $0.12$~dex than an average MW mass system, see Sec.~\ref{sec:radial_profiles}. However, the median $r_{\rm peri}$ of the MW overlaps the $16-23\%$ (within $16-84\%$) levels of the model predictions, meaning that the model can still match this property of the MW GCs. Although the model tends to over-predict the pericenter and apocenter radii for the MW GC system, it can match the observed orbital eccentricity, which is defined as $e = (r_{\rm apo}-r_{\rm peri})/(r_{\rm apo}+r_{\rm peri})$. We find the median eccentricity of the model to be $e=0.62\pm0.04$ for MW mass galaxies, in good agreement with the observed value $e = 0.60\pm 0.02$.

We also list in Table~\ref{tab:properties} the median actions of the 13 modeled MW mass galaxies and the MW. The model can match the three observed median actions of the MW GC system as the radial, azimuthal, and vertical median actions of the MW intersect with the $0-38\%$, $23-47\%$, and $15-41\%$ values of model predictions. However, the median radial action $J_{\rm R}$ of the MW has large uncertainties, with a lower boundary smaller than the median $J_{\rm R}$ of any of the 13 modeled systems. This is likely due to the model's tendency to overestimate the radii of GCs, enlarging the integral range in Eq.~(\ref{eq:action}) to produce larger $J_R$.

\section{Globular clusters from different origins}
\label{sec:gcs_from_different_origins}

Origins of model GCs can be easily distinguished by looking at their positions at birth in the galaxy merger tree. We define the GCs formed in the main progenitor branch as \textit{in-situ} clusters, and the other GCs originally formed in satellite galaxies as \textit{ex-situ} clusters. The \textit{ex-situ} clusters are later brought into the central galaxy via accretion and mergers. In this section, we compare multiple properties of the model clusters formed \textit{in-situ} and \textit{ex-situ} to those of their observed counterparts. However, since we do not have the actual merger tree for the MW, the classification of MW GCs is not as straightforward as the model GCs. Therefore, we adopt the criteria of \citet{massari_origin_2019} to classify the \textit{in-situ} and \textit{ex-situ} components of the MW GCs. These authors define GCs with apocenter radius less than 3.5 kpc as \textit{bulge} clusters, and GCs with maximum height from the disc less than 5 kpc and orbit circularity greater than 0.5 as \textit{disk} clusters. The bulge and disk clusters combined are the \textit{in-situ} GC population, while the rest are \textit{ex-situ} clusters. This classification still has limitations and cannot be regarded as the actual origins of MW GCs. We will mention these caveats and their effects on various GC properties in this section.

\begin{figure}
	\includegraphics[width=\linewidth]{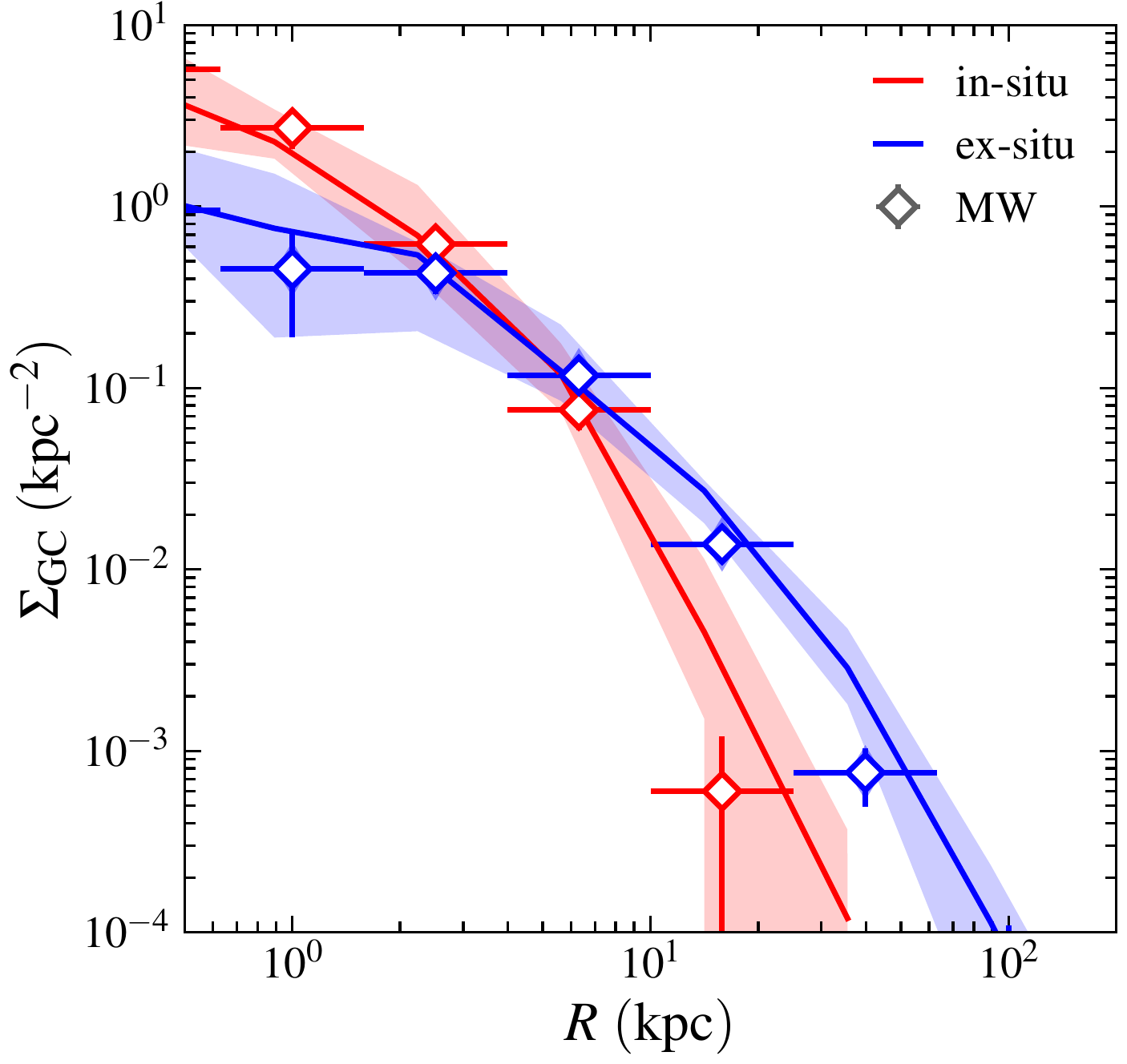}
	\vspace{-5mm}
    \caption{Surface number density profile of modeled GC systems in MW mass galaxies, for the case $\Omega_{\rm tid}=\Omega_{\lambda}$. The \textit{in-situ} and \textit{ex-situ} GCs are plotted as red and blue curves, respectively. The $16-84$th percentiles of each component are shown as shaded regions. For comparison with observations, we plot the dispersion profile of the \textit{in-situ} and \textit{ex-situ} GCs from MW as diamonds with errorbars: vertical errorbars show the Poisson error, and horizontal errorbars correspond to the bin width.}
    \label{fig:Sigma_r_log_origins}
\end{figure}

In Fig.~\ref{fig:Sigma_r_log_origins} we compare the radial profiles of \textit{in-situ} and \textit{ex-situ} GCs in MW mass galaxies. For both samples the surface densities decrease sharply with galactocentric radius in outer regions, $R\gtrsim 10\ {\rm kpc}$. Some \textit{ex-situ} GCs can be found as far as $100\ {\rm kpc}$ from the galaxy center. In the inner $3\ {\rm kpc}$ the \textit{in-situ} component dominates, while the \textit{ex-situ} profile forms a flat core in the center.

Even though we do not specifically calibrate the merit function on the bimodal split, the profiles of both model \textit{in-situ} and \textit{ex-situ} GCs are consistent with their respective observed counterparts. In Table~\ref{tab:properties}, we present the $16-84\%$ effective radii of two GC populations for the model and observations. Note that the radial distribution of \textit{in-situ} and \textit{ex-situ} GCs cannot be fitted by the de Vaucouleurs law any longer, and therefore we use the face-on projected half number radius as the effective radius, $R_{\rm e}$. The model predicts the ratio $R_{\rm e,in}/R_{\rm e}=0.35-0.99$ ($16-84$th percentiles), indicating that the \textit{in-situ} component is systematically more centrally concentrated than the whole GC system. The observed ratio ranges between $0.36-0.58$, overlapping the $17-49$th percentiles of the model prediction. In addition, the model yields $R_{\rm e,ex}/R_{\rm e}=1.36-2.69$, while the observed ratio ($1.38-2.14$) overlaps the $20-68$th percentiles of the model prediction. Although the model results are statistically consistent with observations, the model tends to predict systematically larger $R_{\rm e}$ for both \textit{in-situ} and \textit{ex-situ} GCs since the MW GC system is more compact than average GC systems in the observational samples, as shown in Sec.~\ref{sec:radial_profiles}. Nevertheless, the observed effective radii still overlap the $10-21\%$ (\textit{in-situ}) and $9-45\%$ (\textit{ex-situ}) values of model predictions, indicating that the size of the MW GC system is below average but still typical. 

\begin{figure}
	\includegraphics[width=\linewidth]{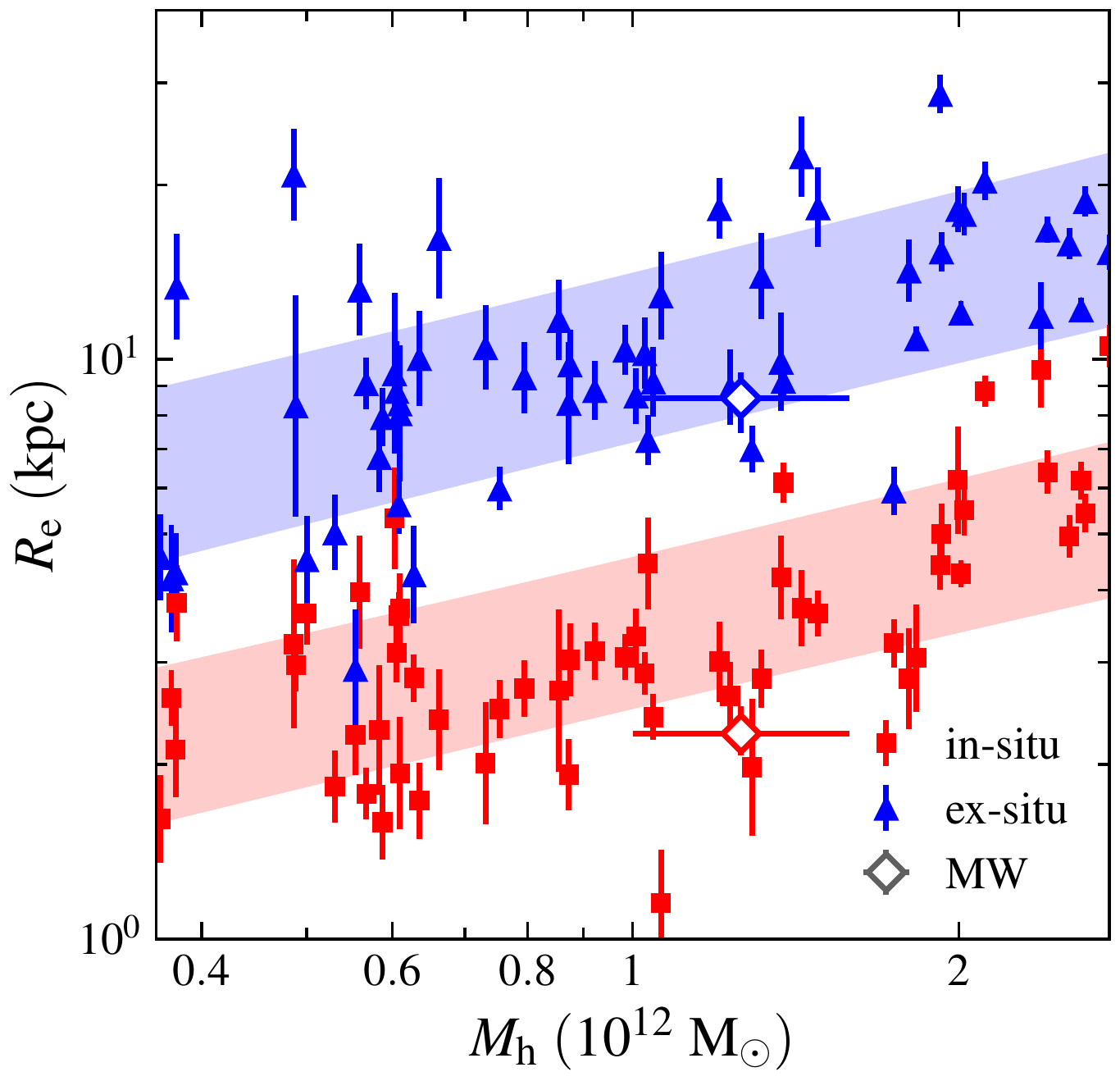}
	\vspace{-4mm}
    \caption{Effective radii of \textit{in-situ} (red squares) and \textit{ex-situ} (blue triangles) components of modeled GC systems. Similarly to Fig.~\ref{fig:re_mh_log}, we show power-law fits of the $R_{\rm e}$--$\Mh$ relations as shaded regions. The \textit{in-situ} (red) and \textit{ex-situ} (blue) components of the MW GC system are plotted as diamonds with errorbars: the horizontal errorbars show the uncertainty of the total mass of MW, and the vertical errorbars correspond to the uncertainties of effective radii from bootstrap resampling.}
    \label{fig:re_mh_log_in_and_ex}
\end{figure}

We show the effective radii of \textit{in-situ} and \textit{ex-situ} GCs for all model galaxies as a function of host halo mass in Fig.~\ref{fig:re_mh_log_in_and_ex}. Similarly to the previous analysis of the $R_{\rm e}$--$M_{\rm h}$ relation for all GCs, the effective radii of both \textit{in-situ} and \textit{ex-situ} GCs scale as power-law functions of host halo mass, with power-law indices of $0.44\pm0.07$ and $0.46\pm0.08$ (obtained from standard linear fit). The intrinsic scatters of the \textit{in-situ} and \textit{ex-situ} components are $0.13\pm0.02$~dex and $0.15\pm0.02$~dex, respectively. Moreover, the effective radii of the \textit{ex-situ} GCs are greater than the \textit{in-situ} ones by $\sim$0.5 dex, which is significantly greater than their intrinsic scatter. The significant discrepancy indicates that \textit{in-situ} and \textit{ex-situ} GCs are distributed at distinct regions. However, since the radial spreads are large for both populations, it is still a big challenge to distinguish the progenitors of GCs by looking merely at the radius.

\begin{figure}
	\includegraphics[width=\linewidth]{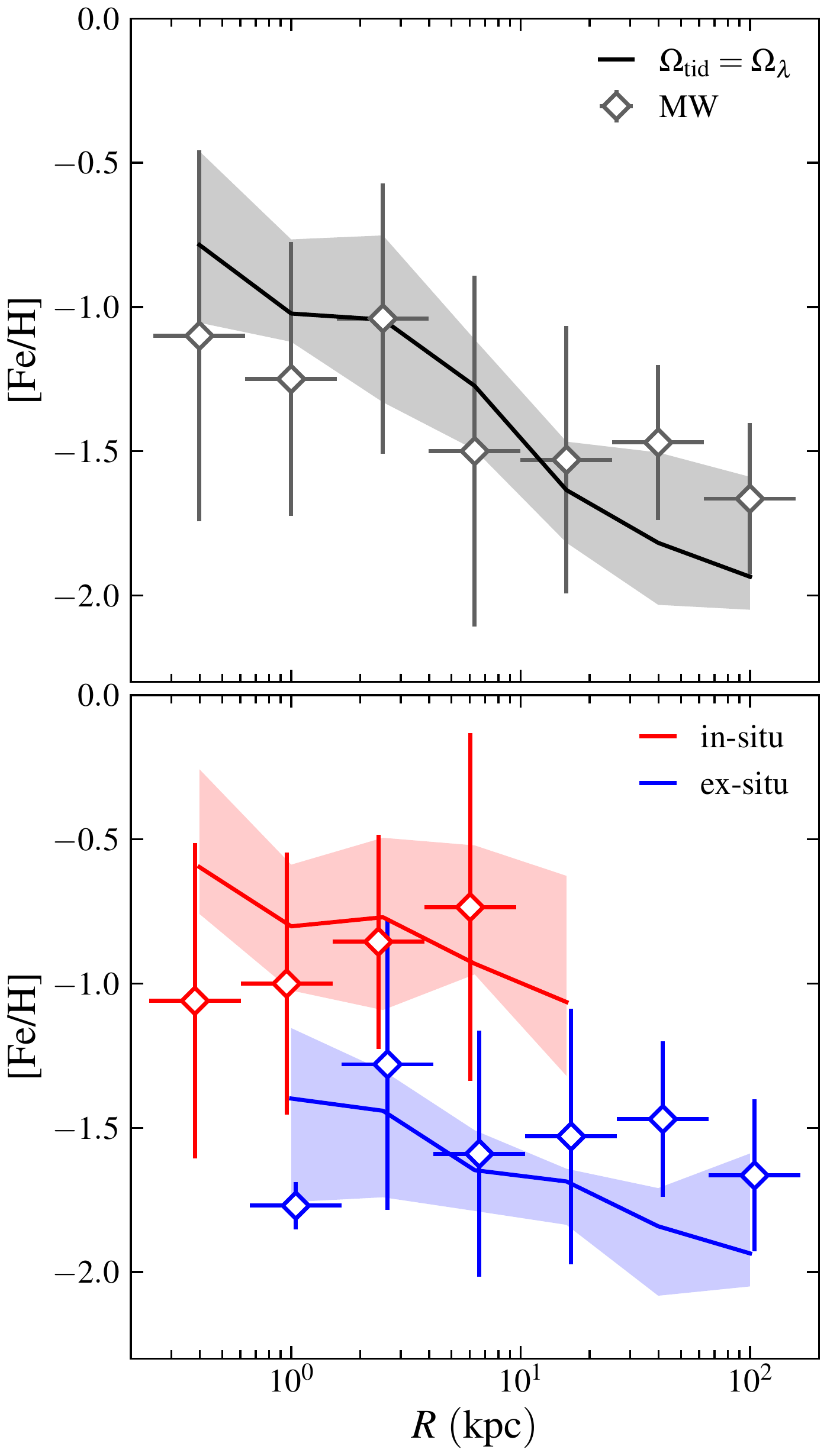}
 	\vspace{-4mm}
    \caption{Radial profiles of metallicity, $\feh$, for modeled GC systems in MW mass galaxies (\textit{top panel}) and for \textit{in-situ} (red) and \textit{ex-situ} (blue) GCs (\textit{bottom panel}). Shaded regions represent the $16-84$th percentiles of each component. We also show the metallicity profiles for the MW GC system as diamonds with errorbars: vertical errorbars represent the $16-84$th percentiles of $\feh$ in each bin, and the horizontal errorbars correspond to the bin width.}
     \label{fig:feh_r_log_origins}
\end{figure}

The significant different radial distributions of \textit{in-situ} and \textit{ex-situ} GCs lead to an interesting phenomenon: the average GC metallicity decreases with radius. We plot the radial profile of metallicity, $\feh$, for the model GC systems in MW mass galaxies in the top panel of Fig.~\ref{fig:feh_r_log_origins}. The metallicity profiles for the \textit{in-situ} and \textit{ex-situ} clusters are shown in the bottom panel of Fig.~\ref{fig:feh_r_log_origins}. For comparison, we also plot the $\feh$ profiles for the MW GCs provided by \citet[][2010 edition]{harris_catalog_1996}. The modeled \textit{ex-situ} GCs are systematically more metal-poor than the \textit{in-situ} GCs by 0.5 to 1 dex. This is because \textit{ex-situ} GCs are more likely to be formed in older and smaller galaxies, where the metallicity is significantly lower than in the main progenitor galaxy (see Eq.~\ref{eq:metalicity_stellar_mass}). In addition, we note that there is no clear dependence on radius for metallicities of both \textit{in-situ} and \textit{ex-situ} GCs, in agreement with the flat metallicity profile in the outer MW halo \citep{searle_compositions_1978}. However, the average metallicity of all GCs drops significantly with radius because the proportion of \textit{ex-situ} GCs grows at large radii. A similar trend also exists in the MW GC system but is obscured by the relatively large scatter. It is reasonable to suggest that the metallicity of GCs in a given radial range can be viewed as a tracer of the abundance of \textit{in-situ} vs \textit{ex-situ} GCs.

\begin{figure*}
	\includegraphics[width=0.9\linewidth]{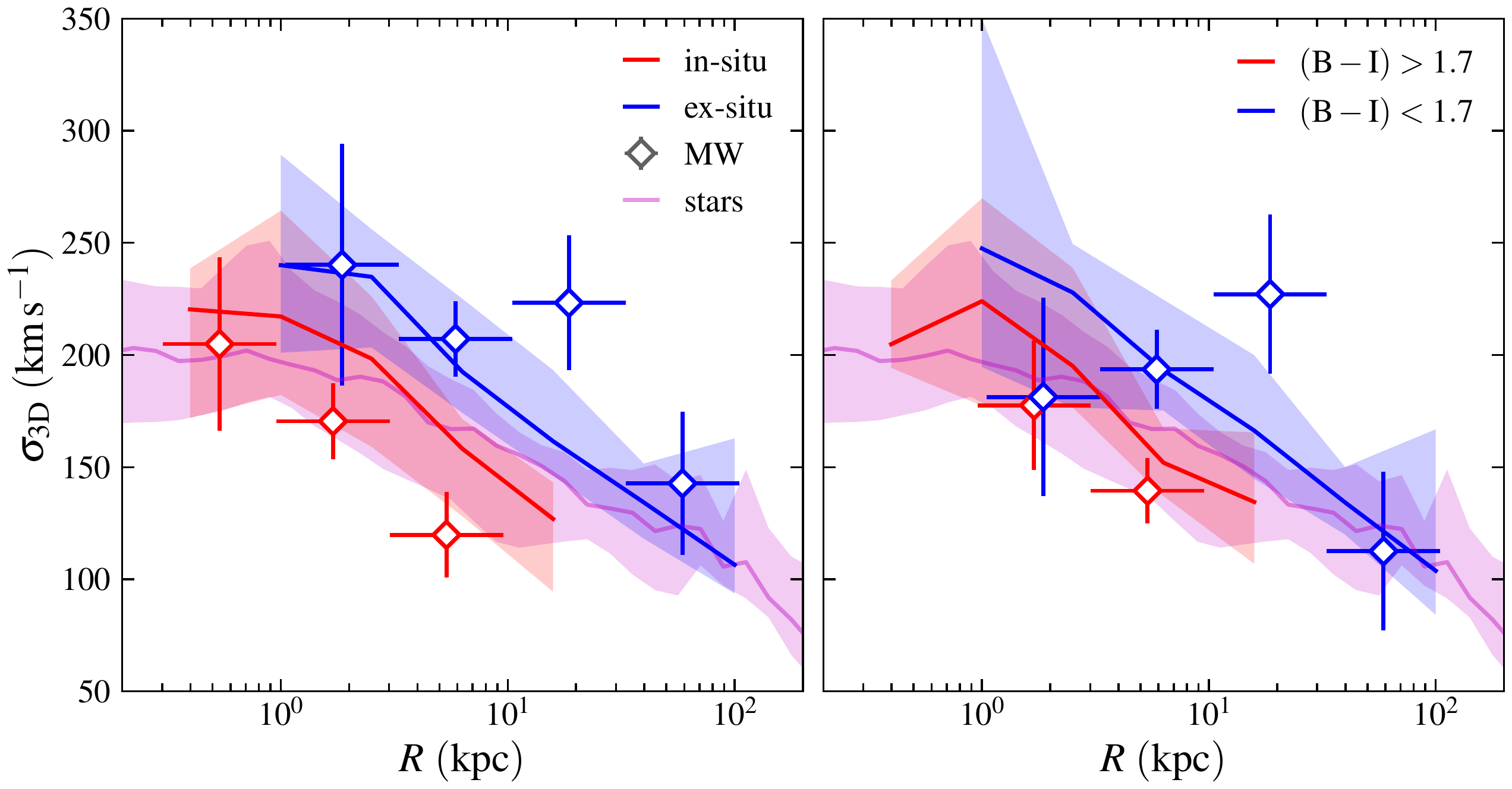}
	\vspace{-1mm}
    \caption{Radial profiles of 3D velocity dispersion for different components of GCs in MW mass galaxies. In the \textit{left panel}, GCs are split into \textit{in-situ} (red) and \textit{ex-situ} (blue) components; whereas GCs are split as red (${\rm (B-I)}>1.7$) and blue (${\rm (B-I)}<1.7$) components in the \textit{right panel}. Other parameters are similar to those in Fig.~\ref{fig:dispersion_r}.}
    \label{fig:dispersion_r_origins}
\end{figure*}

Next, we investigate the radial profiles of 3D velocity dispersions for \textit{in-situ} and \textit{ex-situ} clusters. Fig.~\ref{fig:dispersion_r_origins} shows that both samples have $\sigma_{\rm 3D}$ decreasing with radius. Although the trend can be easily noticed by looking at the average $\sigma_{\rm 3D}$ profile, the profiles for individual galaxies can greatly deviate from the average as the intrinsic scatter can be as large as $30-50\kms$. This is also true for the MW GCs. Although the predicted dispersion profiles are mostly consistent with observations at the $16-84\%$ confidence level, the observed dispersion profile of \textit{ex-situ} clusters peaks dramatically at $R\simeq30$ kpc. As discussed in Sec.~\ref{sec:kinematics}, this bump is likely due to the high velocities of Sagittarius stream clusters. Moreover, the $\sigma_{\rm 3D}$ for \textit{in-situ} GCs is systematically lower than the \textit{ex-situ} ones by $\sim$40$\kms$ in the range where they overlap, $R=1-20$ kpc. The higher dispersion for \textit{ex-situ} clusters is likely because the \textit{ex-situ} GCs come from several satellite galaxies with distinct kinematics, and many of them are brought into the main progenitor galaxy via violent gravitational interactions, leading to greater velocity dispersion. The migration nature of the modeled \textit{ex-situ} clusters also results in their velocity dispersions being systematically larger than the field stars at $R=1-20$ kpc, whereas the modeled \textit{in-situ} clusters have velocity dispersions similar to that of the field stellar component. The deviation between \textit{in-situ} and \textit{ex-situ} clusters is even larger in the MW as the observed \textit{in-situ} clusters have lower dispersions than the model median. This is likely because the classification of MW GCs is based on \citet{massari_origin_2019}, in which \textit{in-situ} GCs are arbitrarily defined as clusters with low $r_{\rm apo}$ (a.k.a. bulge clusters) or high circularity (a.k.a. disk clusters). These criteria favor GCs with greater bulk rotational velocities rather than random motions, leading to the selected \textit{in-situ} clusters having lower dispersions.

Observationally, we cannot easily distinguish the origins of extragalactic GCs. It is therefore more applicable to compare the dispersion of GCs split by the color index instead of the \textit{in-situ} vs \textit{ex-situ} origins. Following \citet{harris_globular_2006}, we compute the metallicity sensitive ${\rm (B-I)}$ color via a linear relation:
\begin{equation}
	{\rm (B-I)}=2.158+0.375\,\feh.
	\label{eq:feh2B-I}
\end{equation}
Since the metallicity of GCs can serve as a tracer of \textit{in-situ} vs \textit{ex-situ} GCs, Eq.~(\ref{eq:feh2B-I}) indicates that the ${\rm (B-I)}$ color can also trace the GC origins. For the modeled MW mass galaxies, we find the median GC ${\rm (B-I)}$ varying between $1.5$ and $1.9$. Without loss of generality, we show the 3D velocity dispersions of red and blue GCs split at ${\rm (B-I)}=1.7$ in the right panel of Fig.~\ref{fig:dispersion_r_origins}. Again, the 3D velocity dispersions decrease with radius. Compared with blue GCs, red GCs have $\sim40\kms$ lower velocity dispersion. The red clusters have velocity dispersions similar to the stellar components, while the blue clusters have velocity dispersions systematically larger than the field stars at the same radius. Although the MW GC system does not show exactly the same behaviour as the modeled trends, we still observe the blue GCs in the MW to have greater dispersion than the red ones at $R\simeq 10\ {\rm kpc}$. Velocity dispersion differences between the red and blue GCs are observed more clearly in giant elliptical galaxies, such as NGC 1399 \citep[see Fig.~14 of][]{schuberth_globular_2010}. It is thus reasonable to suggest that the different origins of GCs can contribute to the observed dispersion difference between the red and blue components. 

\begin{figure}
	\includegraphics[width=\linewidth]{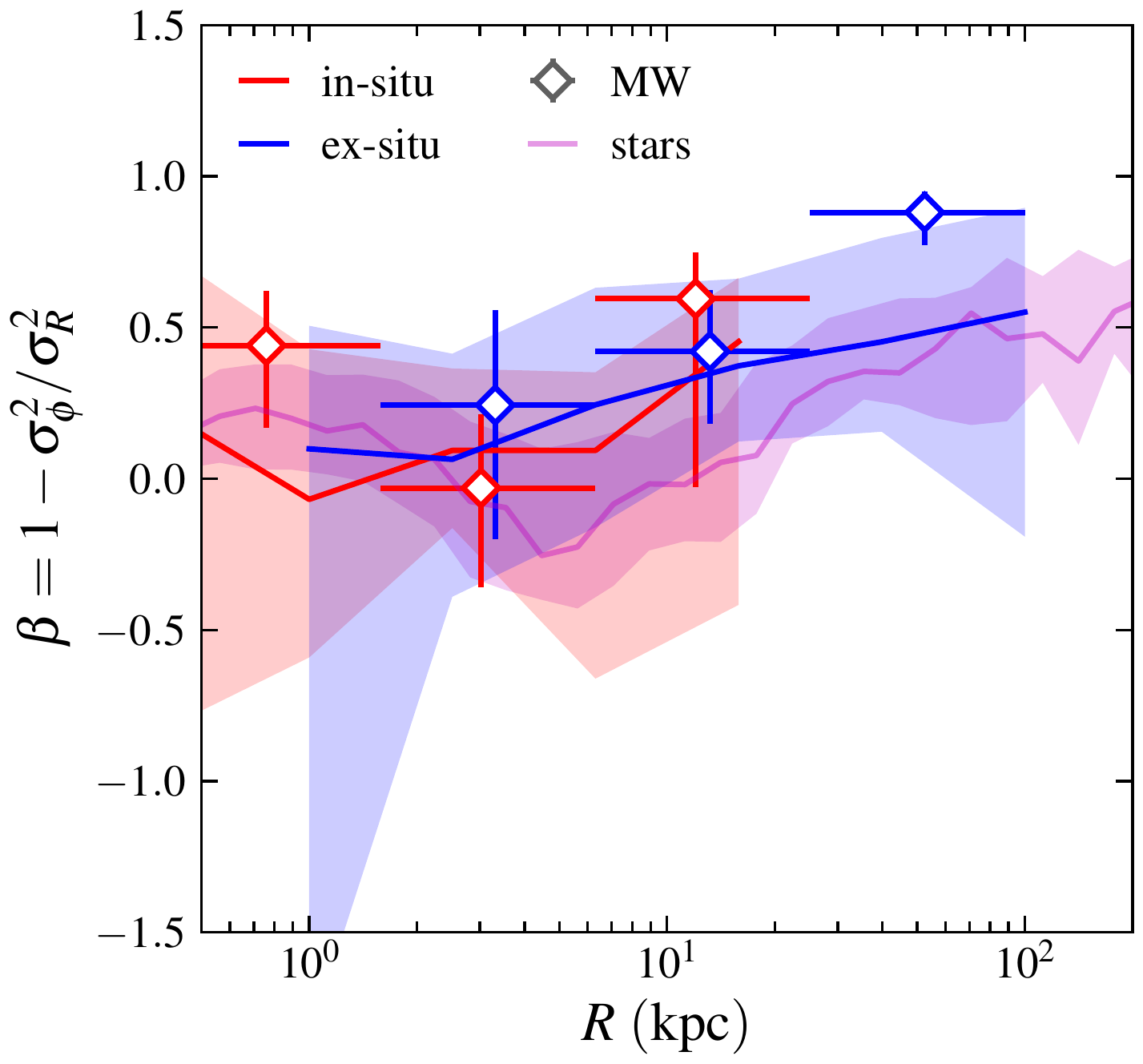}
 	\vspace{-5mm}
    \caption{Radial profiles of anisotropy parameter for \textit{in-situ} (red) and \textit{ex-situ} (blue) GCs (\textit{bottom panel}) in MW mass galaxies. Other parameters are similar to those of Fig.~\ref{fig:beta_r_log}.}
     \label{fig:beta_r_log_origins}
\end{figure}

We also study the radial profiles of the anisotropy parameter for \textit{in-situ} and \textit{ex-situ} clusters. As shown in Fig.~\ref{fig:beta_r_log_origins}, both GC populations have the anisotropy parameter consistent with zero at $R = 1-10$ kpc, similarly to field stellar particles in the same radial range. At $R > 10$ kpc, where \textit{ex-situ} GCs dominate, we find that the anisotropy parameter of \textit{ex-situ} clusters is systematically positive, suggesting that the motions of outer GCs are radially dominated, in agreement with the accretion nature of \textit{ex-situ} GCs. The model predictions of \textit{in-situ} and \textit{ex-situ} GCs are consistent with the observations within the $16-84\%$ confidence level.

Finally, we calculate the median pericenter/apocenter radii and orbital actions for \textit{in-situ} and \textit{ex-situ} GCs in 13 MW mass galaxies. Similarly to the way we present data for all GCs, we present in Table~\ref{tab:properties} these properties for the modeled systems, the MW, and the respective percentiles. The median pericenter and apocenter radii of the \textit{ex-situ} GCs are greater than those of the \textit{in-situ} ones, consistent with the migration nature of \textit{ex-situ} GCs. For the two median radii, the $16-84\%$ ranges of the two GC populations do not overlap. This again supports that GCs from different progenitors have systematically different radial distributions. The deviations between the two GC systems are especially notable for the median $r_{\rm apo}$, which is 3.5 times higher for the \textit{ex-situ} clusters. However, note that what we present here are the median values. Since the spreads of the these radii in individual galaxies are rather broad, it is still challenging to distinguish the progenitors of each GC by looking only at $r_{\rm peri}$ and $r_{\rm apo}$. 

The observed median $r_{\rm peri, MW}$ overlaps the $15-70\%$ (\textit{in-situ}) and $0-10\%$ (\textit{ex-situ}) values of model predictions. And, the median $r_{\rm apo, MW}$ overlaps the $7-27\%$ (\textit{in-situ}) and $12-52\%$ (\textit{ex-situ}) values of model predictions. Except for the model systematically over-predicting median $r_{\rm peri}$, the model can mostly match the observed values. We also note that the median $r_{\rm apo}$ of the \textit{ex-situ} MW GCs is greater than $r_{\rm apo}$ \textit{in-situ} by a factor of $5$, which is even larger than the predicted difference. However, note that the \citet{massari_origin_2019} classification arbitrarily defines \textit{in-situ} GCs as clusters with low $r_{\rm apo}$ or high circularity. The former criterion increases the discrepancy between $r_{\rm apo}$ of the \textit{in-situ} and \textit{in-situ} GCs. The latter favors GCs with more circular orbits. Therefore, the orbits of remaining \textit{ex-situ} clusters are more eccentric. That is to say, GCs with small $r_{\rm peri}$ and large $r_{\rm apo}$ are more likely to be classified as \textit{ex-situ} clusters. This bias explains why the model over-predicts the median $r_{\rm peri}$ for \textit{ex-situ} GCs. The difference between our results and the results from \citet{massari_origin_2019} suggests that these authors might underestimate the effects of dynamical evolution on the orbits of MW GCs. The actual orbits of \textit{in-situ} GCs may be more eccentric and have higher $r_{\rm apo}$ than expected.

The median orbital actions $(J_R, |J_\phi|, J_z)$ for \textit{in-situ} and \textit{ex-situ} GCs are also presented in Table~\ref{tab:properties}. The difference between the two GC populations is the largest for $J_R$ and the smallest for $|J_\phi|$, indicating that $J_R$ can be a useful parameter to distinguish GCs from different progenitors. Compared with the model predictions, the observed median $J_{R,{\rm MW}}$ overlaps the $0-21\%$ (\textit{in-situ}) and $54-92\%$ (\textit{ex-situ}) values, the median $|J_{\phi,{\rm MW}}|$ overlaps the $22-92\%$ (\textit{in-situ}) and $0-12\%$ (\textit{ex-situ}) values, and the median $J_{z,{\rm MW}}$ overlaps the $16-30\%$ (\textit{in-situ}) and $29-43\%$ (\textit{ex-situ}) values of the 13 modeled MW mass galaxies. Most of the model predictions can match the observed values, except for the overestimation for \textit{in-situ} $J_R$ and \textit{ex-situ} $|J_\phi|$. As discussed before, the overestimation for \textit{in-situ} $J_R$ is because the model \textit{in-situ} GCs have larger $r_{\rm apo}$, enlarging the integral range in Eq.~(\ref{eq:action}) to produce larger $J_R$. Also, since the \citet{massari_origin_2019} classification tends to select \textit{ex-situ} GCs with more eccentric orbits, their \textit{ex-situ} GCs are likely to have lower $|L_z|$ (recall that $J_\phi=L_z$) than reality. Consequently, our model over-predicts the median $|J_\phi|$ for \textit{ex-situ} GCs. 

\section{Discussion}
\label{sec:discussion}

\subsection{Comparison with cases of constant disruption rate}
\label{sec:comparison_with_constant_disruption_rate}

\begin{figure}
	\includegraphics[width=\linewidth]{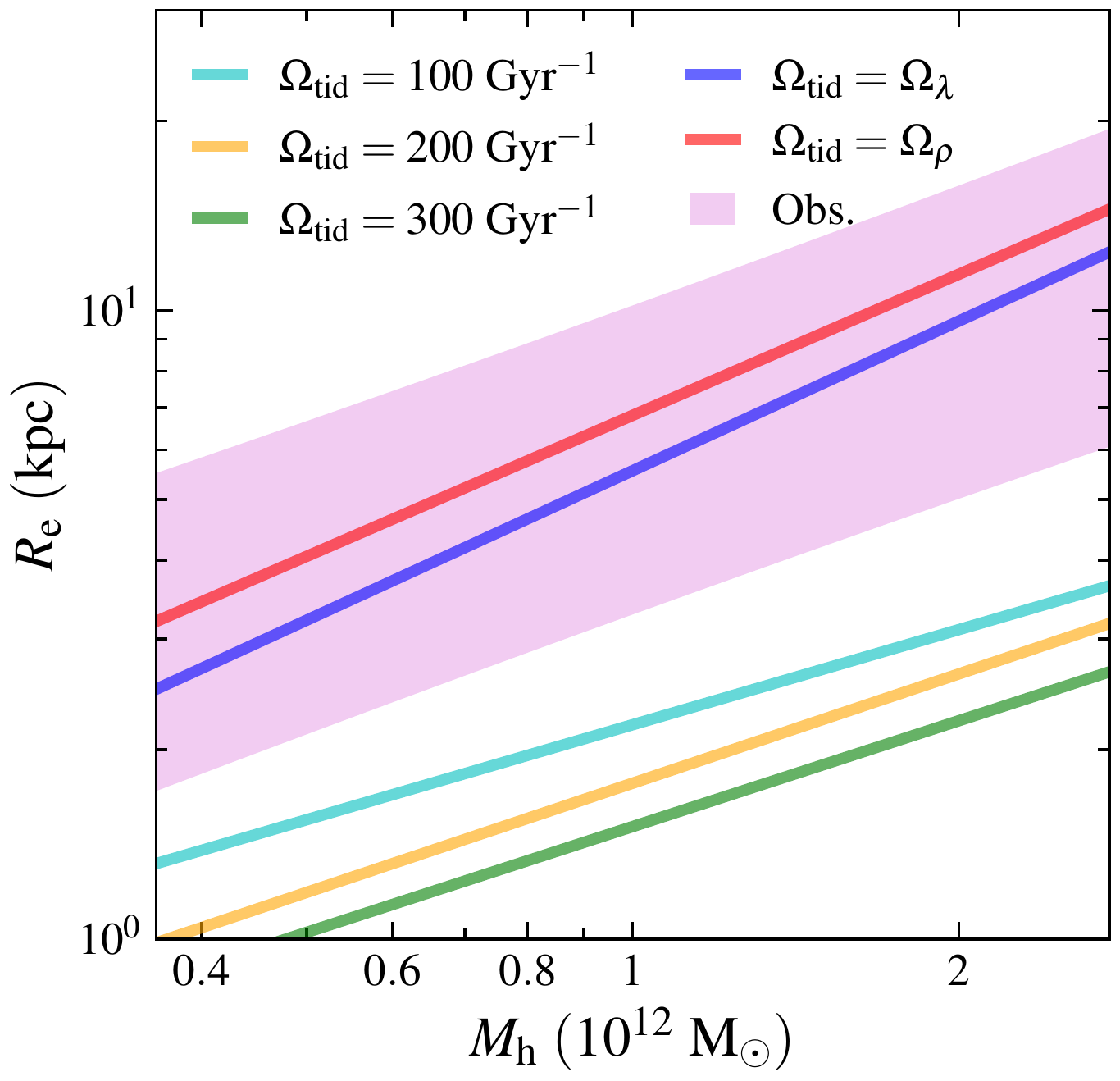}
	\vspace{-5mm}
    \caption{Best-fit de Vaucouleurs effective radii, $R_{\rm e}$, of modeled GC systems as a function of host halo mass. The $\Omega_{\rm tid}=\Omega_{\lambda}$ and $\Omega_{\rm tid}=\Omega_{\rho}$ cases are shown as blue and red lines, while the constant disruption rate cases of $\Omega_{\rm tid}=100$, $200$, and $300\ {\rm Gyr^{-1}}$ are shown as cyan, orange, and green lines, respectively. Other parameters are as in Fig.~\ref{fig:re_mh_log}.}
    \label{fig:re_mh_log_with_constant}
\end{figure}

In Sec.~\ref{sec:radial_profiles} we note that the effective radii of model GC systems can be described as a power-law function of galaxy mass: $R_{\rm e}\propto M_{\rm h}^{0.79\pm 0.09}$. Here, we investigate the effects of different prescriptions for tidal disruption on the $R_{\rm e}$--$M_{\rm h}$ relation. In addition to the $\Omega_{\rm tid}=\Omega_\lambda$ and $\Omega_{\rm tid}=\Omega_\rho$ cases, we introduce another prescription with $\Omega_{\rm tid}=$ constant, which is employed by our previous models \citep{choksi_origins_2019}. Due to the lack of spatial information, \citet{choksi_origins_2019} simply set $\Omega_{\rm tid}=200\ {\rm Gyr^{-1}}$ for all clusters at all times.\footnote{In their notation, $P=0.5$, where $P=\left(\frac{\Omega_{\rm tid}}{100\ {\rm Gyr^{-1}}}\right)^{-1}$.} Taking this setup as a reference, we examine the values $\Omega_{\rm tid}=100$, $200$, and $300\ {\rm Gyr^{-1}}$ for completeness and compare these models with the $\Omega_{\rm tid}=\Omega_\lambda$ and $\Omega_{\rm tid}=\Omega_\rho$ cases. Note that $\kappa$ is no longer a model parameter since $\Omega_{\rm tid}$ is now fixed. By performing the same calibration as in Sec.~\ref{sec:model_calibration} to search for the two remaining model parameters, we find $p_2=6$, $12$, and $20$ for $\Omega_{\rm tid}=100$, $200$, and $300\ {\rm Gyr^{-1}}$, respectively; while $p_3=0.6\ {\rm Gyr^{-1}}$ works well for all of them. The $R_{\rm e}$--$M_{\rm h}$ relations for these cases are shown in Fig.~\ref{fig:re_mh_log_with_constant}. Compared with the disruption prescriptions employed in this work, the $R_{\rm e}$--$M_{\rm h}$ relations of all constant $\Omega_{\rm tid}$ cases have flatter slopes ranging from $0.5$ to $0.6$, which are flatter than the $0.7-0.8$ slopes from the $\Omega_{\rm tid}=\Omega_{\lambda/\rho}$ cases. Compared with observations, the three $\Omega_{\rm tid}=$ constant cases also agree with the $0.62\pm 0.13$ slope within the error range. However, the three constant $\Omega_{\rm tid}$ cases have significantly smaller normalization by a factor of $\sim3$ compared to the $\Omega_{\rm tid}=\Omega_{\lambda/\rho}$ cases and observations.

The present-day radial distribution of a GC system can significantly differ from the initial distribution. For example, systemic radial motions can bring GCs inwards or outwards, shifting the radial distribution from the initial one. Different dependence of disruption on local environment can also lead to different radial distribution of surviving GCs. Compared with the environment-dependent models ($\Omega_{\rm tid}=\Omega_{\lambda/\rho}$), the $\Omega_{\rm tid}=$ constant cases tend to produce stronger tidal disruption in the outer galaxy since they do not take into account the strength of tidal disruption decreasing with radius (see Appendix~\ref{sec:accuracy_of_approximating_tidal_disruption} for more details of the decreasing trend). This directly leads to higher disruption rate in the outer galaxy for \textit{in-situ} clusters. On the other hand, the \textit{ex-situ} GCs also experience systematically stronger tidal disruption in the outer region after merging with the central galaxy. Therefore, the $\Omega_{\rm tid}=$ constant cases have a tendency to form more centrally concentrated GC systems. We also note that the $\Omega=\Omega_\rho$ case produces generally larger $R_{\rm e}$ than the $\Omega=\Omega_\lambda$ case. This is due to the larger disruption rate at small radii for the $\Omega=\Omega_\rho$ case (see Appendix~\ref{sec:accuracy_of_approximating_tidal_disruption}). Since more central GCs are disrupted, the $\Omega=\Omega_\rho$ case has a tendency to form more spread-out GC systems than the $\Omega_{\rm tid}=\Omega_\lambda$ case.

Nevertheless, we cannot assert either of the $\Omega_{\rm tid}=\Omega_{\lambda/\rho}$ cases to be more appropriate than the constant disruption cases, since the initial spatial distribution of GCs in our model is not guaranteed to be correct. A more dispersed initial radial distribution may raise the effective radius to a more reasonable range for the $\Omega_{\rm tid}=$ constant cases. 

\subsection{GC mass function}
\label{sec:gc_mass_function}

\begin{figure}
	\includegraphics[width=\linewidth]{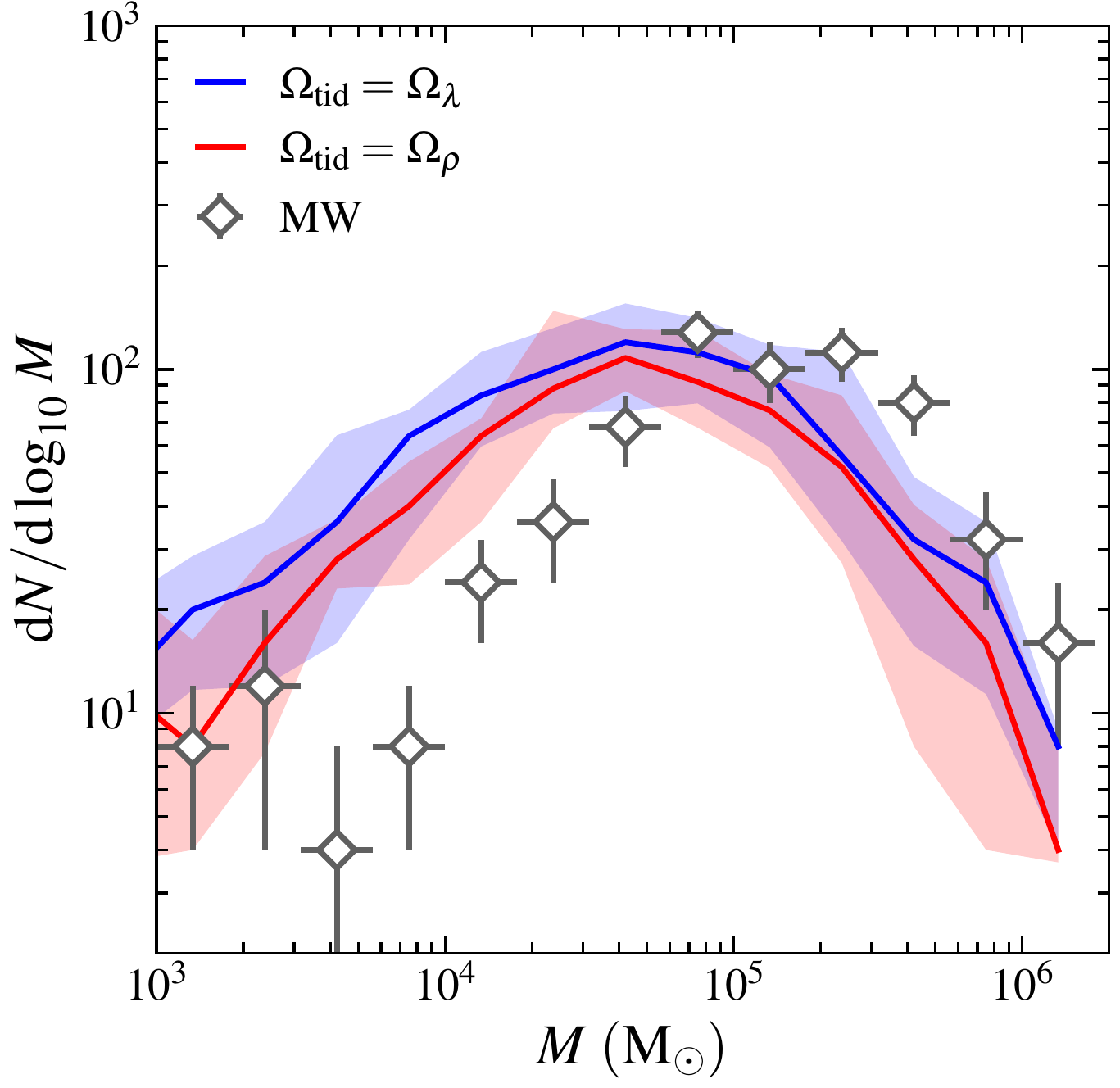}
	\vspace{-4mm}
    \caption{GC mass function of GC systems in MW mass galaxies. The two descriptions of tidal disruption, $\Omega_{\rm tid}=\Omega_{\lambda}$ and $\Omega_{\rm tid}=\Omega_{\rho}$, are shown as blue and red curves, with shaded regions representing the $16-84$th percentiles of the $13$ galaxies. For comparison, the mass function of the MW GC system is overplotted as diamonds with errorbars: vertical errorbars show the $16-84\%$ confidence level computed via bootstrap resampling, and horizontal errorbars correspond to the bin width. We repeat bootstrap resampling 1000 times until the estimated confidence levels converge.}
    \label{fig:mass_function}
\end{figure}

Starting with a Schechter ICMF, the tidal and stellar evolution of GCs reshapes the mass distribution into the present-day GC mass function (GCMF). In our model, stellar evolution is included as an instantaneous mass loss after the formation of GCs. Since the fraction of mass lost by stellar evolution is independent of GC mass (see Eq.~\ref{eq:stellar_evolution}), only tidal disruption is important for transforming the shape of ICMF to GCMF. According to Eq.~(\ref{eq:t_tid}), the high-mass end of the ICMF is less affected by tidal disruption and can preserve its initial shape, while the low-mass end can turn over because of the very efficient disruption of small GCs. In Fig.~\ref{fig:mass_function} we show that the GCMFs predicted by our model for the MW mass galaxies have similar shape and normalization to the observed MW GCMF. Our GCMFs also agree with \citet{hughes_physics_2022}, who conducted a study on the high-mass end GCMFs in the E-MOSAICS simulation. They suggested that the high-mass end of GCMF preserves the initial Schechter shape, with a truncation mass of $\sim 10^6\Msun$ (for MW mass galaxies). Our model can produce similar results when following the same analysis as in their work. On the other hand, there is a small deviation between the peaks of the modeled and observed GCMF: the model GCMF peaks at $M=10^{4.5}-10^{5}\Msun$, whereas the MW GCMF peaks at $M=10^{5}-10^{5.5}\Msun$. Therefore, the model tends to overestimate the number of $M\lesssim10^5\Msun$ GCs and underestimate the number of GCs with higher mass. 

It may be expected that this discrepancy could be resolved by adjusting the three model parameters (Sec.~\ref{sec:modeling_cluster_formation_and_evolution}), as the combination of $p_2$ and $p_3$ controls the total number of GCs formed in the model, i.e., the normalization of the ICMF; and $\kappa$ controls the strength of disruption, which bends the low mass tail of the ICMF. However, we find that even though increasing $\kappa$ can shift the peak mass of GCMF to higher values, the change in peak mass is small compared to the increase of $\kappa$. Setting $\kappa=10$ still cannot produce a GCMF matching the observations at the low-mass end, but can significantly affect the radial distribution of GCs by disrupting too many inner GCs. Since the low-mass GCs cannot be effectively disrupted with the current tidal disruption prescription, a more realistic prescription is needed to model the mass loss of $M\lesssim10^5 {\Msun}$ GCs due to tidal shocks.

\subsection{How GCs migrate from the current locations}
\label{sec:migration}

\begin{figure}
 	\includegraphics[width=\columnwidth]{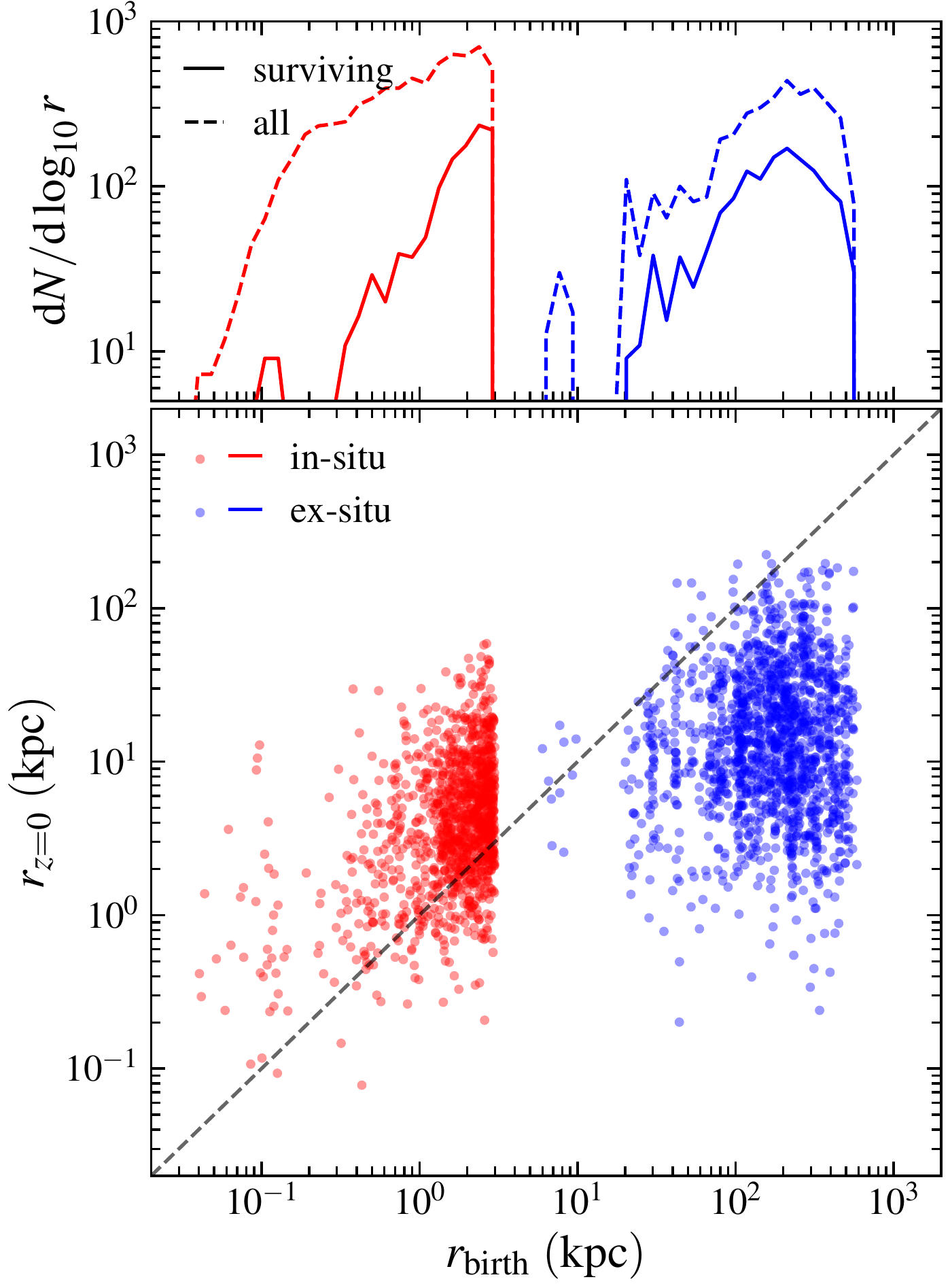}
 	\vspace{-4mm}
    \caption{Comparison between radii at $z=0$ and at formation for \textit{in-situ} (red) and \textit{ex-situ} (blue) GCs in the 13 MW mass galaxies (\textit{bottom panel}). The gray line represents $r_{z=0}=r_{\rm birth}$. We plot the average radial distribution of surviving (solid) and all (i.e., surviving $+$ disrupted, dashed) GCs at formation in the \textit{top panel}. }
    \label{fig:r_z0_r_birth}
\end{figure}

The present-day distribution of GCs is shaped by the interplay of initial distribution, dynamical disruption, and migration. Here, we investigate how GCs migrate from the original positions to the current ones, and the different disruption in the inner/outer parts of galaxies. In Fig.~\ref{fig:r_z0_r_birth}, we compare the present-day radii to the radii at formation for all GCs from the 13 MW mass galaxies. Since the orientation of the galaxy plane is not well defined at the early time when GC form, here we use spherical radii $r$ instead of cylindrical radii $R$, which we use for the rest of this work. The \textit{ex-situ} GCs form mostly at $r_{\rm birth}=20-500\ {\rm kpc}$ from the galaxy center, while the \textit{in-situ} clusters form in the inner region $r_{\rm birth}=0.1-3\ {\rm kpc}$. The hard cut for \textit{in-situ} clusters at $3\ {\rm kpc}$ is imposed in Sec.~\ref{sec:cluster_sampling}.

Most \textit{in-situ} clusters migrate outwards to as far as $r_{z=0}\simeq100\ {\rm kpc}$, whereas \textit{ex-situ} clusters get accreted by the main progenitor galaxy and move inwards. Although formed in distinct regions, both GC populations relocate to similar present-day regions between $r_{z=0}=1$ and 100~kpc. By comparing the distribution of surviving and all (i.e., surviving $+$ disrupted) GCs in Fig.~\ref{fig:r_z0_r_birth}, we also note that the most efficient disruption of \textit{in-situ} clusters happens in inner regions where the tidal field is stronger.

\subsection{Comparison with other work}
\label{sec:comparison_with_other_works}

First, we compare our numerical setup and results with the E-MOSAICS project \citep{pfeffer_e-mosaics_2018,kruijssen_e-mosaics_2019}. The E-MOSAICS project re-simulated the EAGLE \citep{schaye_eagle_2015} suite of galaxy simulations with the MOSAICS \citep{kruijssen_photometric_2008,kruijssen_evolution_2009,kruijssen_modelling_2011} star cluster formation and evolution model. This model treats star clusters as a sub-grid component of stellar particles. When a stellar particle is formed, the MOSAICS model turns a fraction of its mass into star clusters with a cluster formation efficiency based on the local gas density, velocity dispersion, and the sound speed of the cold interstellar medium. In contrast, in our model, the masses of newly formed clusters are determined only by the global properties of the host galaxy. Similar to the particle assignment technique in our work, star clusters in the E-MOSAICS project inherit the spatial and kinematic information from the simulation particles. They also inherit metallicity from the stellar particles, while in our model cluster metallicity is set by the global galaxy metallicity. The E-MOSAICS project requires a very fine time resolution to identify gravitational tidal shocks, which would result in over $10^4$ output snapshots with a correspondingly large amount of data storage. To reduce the storage requirement, the E-MOSAICS project applies the cluster model in the runtime of the simulation.

Using the E-MOSAICS results, \citet{trujillo-gomez_kinematics_2021} analyzed the kinematics of modeled GCs in $25$ MW mass galaxies. They found that the radial velocity dispersion of GC systems is systematically larger than the tangential components, leading to a positive anisotropy parameter, which agrees with our conclusion that the GC orbits are more radially biased. However, most ($>84\%$) GC systems in their work have lower systemic velocity dispersions than the observed values of the MW GCs. They therefore argue that MW is atypical compared to the E-MOSAICS galaxies. We find instead that the observed median dispersions overlap the $41-80\%$ ($\sigma_R$), $64-98\%$ ($\sigma_\phi$), and $29-59\%$ ($\sigma_z$) values of the modeled MW mass galaxies, meaning that our sample can better reproduce the kinematic properties of the MW GC system. In addition, they investigated the radial distribution of velocity dispersions and found that the dispersions are almost flat, whereas we discovered the velocity dispersion to decrease significantly with radius. In both studies, the scatter of the dispersion profile is large, indicating that it is hard to find regularity for a single MW mass galaxy since the formation history of galaxies can be vastly different. For the same reason, it is hard to compare the observed dispersion profile in the MW to either of the two studies. Moreover, \citet{trujillo-gomez_kinematics_2021} computed the median apocenter/pericenter radii of GC orbits and found their results consistent with the observed properties of the MW. In contrast, our model tend to over-predict the median radii since our model is calibrated with a large observational sample set of GC systems, where the MW GC system is by $0.12$~dex more compact than the average MW mass systems. These authors also conducted analysis on different kinematics of \textit{in-situ} and \textit{ex-situ}
clusters. They found \textit{ex-situ} GCs have a stronger tendency to show greater radial velocity dispersion, whereas the tangential dispersions of \textit{in-situ} and \textit{ex-situ} clusters are similar. In combination, the \textit{ex-situ} GCs have greater total velocity dispersion, which agrees with our findings. On the other hand, by splitting GCs into metal-rich and metal-poor at $\feh=-1.2$, they found metal-poor GCs to have greater systemic velocity dispersion. Similar results are also present in our work when splitting by color at $\rm (B-I)=1.7$, corresponding to $\feh\approx-1.2$ by Eq.~(\ref{eq:feh2B-I}).

Also based on the E-MOSAICS results, \citet{reina-campos_globular_2021} investigated the morphology of GC systems as tracers of host galaxies and dark matter halos. By fitting the de Vaucouleurs profile to the projected GC number density, they discovered a strong positive correlation between the effective radii $R_{\rm e}$ and the stellar mass of the host halo, which agrees with our conclusion that $R_{\rm e}$ increases as a power-law function of the halo mass $M_{\rm h}$ (see Fig.~\ref{fig:re_mh_log}). However, these authors tend to overestimate the effective radius for the MW GCs by $\sim0.3$~dex. Although the MW GC system is more compact than an average MW mass system by $\sim0.12$~dex, the $\sim0.3$~dex overestimation in their work is still significant. This overestimation becomes more notable for lower mass galaxies when compared with observations by \citet{forbes_how_2017} and \citet{hudson_correlation_2018}. Since the E-MOSAICS project tends to underestimate tidal disruption, as discussed by \citet{pfeffer_e-mosaics_2018}, the overestimation of $R_{\rm e}$ may be even greater when more realistic disruption is included, as Sec.~\ref{sec:migration} suggests that most disrupted \textit{in-situ} GCs are in the inner galaxy. On the other hand, in general our model predicts lower $R_{\rm e}$, which agrees better with observations. Since the differences between the methods in the E-MOSAICS and this work are many, it is difficult to point exact reason for discrepant $R_{\rm e}$ predictions. One possible reason is that our model applies a radius threshold of $3\ {\rm kpc}$ when assigning GC particles at birth (Sec.~\ref{sec:cluster_sampling}), as recent observations of young clusters \citep{adamo_probing_2015,randriamanakoto_young_2019,adamo_star_2020} suggest that massive clusters preferentially form in the inner regions of galaxies. This threshold also prevents including faraway particles that are not bound to the galaxy. Although the initial distribution is largely modified by the dynamical evolution, the present-day distribution of GCs can still get statistically more centrally concentrated when applying the $3\ {\rm kpc}$ threshold.

We also compare our work with the GC formation and evolution model by \citet{ramos-almendares_simulating_2020}, who used a `GC tagging' technique similar to our GC assignment method. They selected some simulation particles as tracers of GCs based on the merger history of galaxies. In order to be compatible with dark matter-only simulations, their selection criterion is unrelated to any baryonic properties: the tracers are selected to be dark matter particles located within a certain gravitational well. Also, they did not explicitly follow the cluster mass loss due to dynamical disruption. By applying their model to the Illustris simulation \citep{vogelsberger_properties_2014}, these authors performed a detailed analysis of the spatial distribution and kinematics of GC systems. Like this work, they also found that the \textit{in-situ} GCs have more concentrated distribution than the \textit{ex-situ} counterparts, but the distinction between the two populations is relatively small. When split by color, the blue GCs agree with observations, while the modeled red GCs are distributed much more widely than observations. They attributed this deviation to the insufficient intrinsic segregation between different components of GCs. This problem is solved in our work as the predicted radial distributions of both \textit{in-situ} and \textit{ex-situ} GCs are consistent with observations. Additionally, they found that GC systems tend to have positive anisotropy parameter $\beta$, in agreement with our conclusions.

\section{Summary}
\label{sec:summary}

In this work, we present a GC formation and evolution model which explicitly tracks the spatial distribution and kinematics of GC systems. Without running new galaxy formation simulations, we apply the model in post-processing of the TNG50 simulation and select tracers of GCs from collisionless particles according to their age and location. Next, we calculate the mass loss of GCs due to the stellar and tidal evolution, by explicitly taking into account the dependency of tidal disruption on the local environment. The model produces a catalogue of surviving GCs with full spatial and kinematic information. There are only three adjustable parameters in this model, and we calibrate them by comparing the GC catalogue with observations of the MW and a sample of extragalactic GC systems.

Our model succeeds in reproducing important properties of the MW GC system. For example, the radial number density profile in our model matches the observed distribution of MW GCs (Fig.~\ref{fig:Sigma_r_log_mw}). We note that the radial distribution of GCs can be well fit by the de Vaucouleurs law, which is parametrized by the effective radius, $R_{\rm e}$. Our model reveals a power-law scaling relation between $R_{\rm e}$ and the host galaxy mass, in the form $R_{\rm e}\propto M_{\rm h}^{0.79\pm 0.09}$ (Eq.~\ref{eq:simple_fit_model}). The observational measurements of these variables have a very large scatter and not a well-defined slope. Our predicted relation is consistent with the data within the errors (Fig.~\ref{fig:re_mh_log}). We argue that the dependency of tidal disruption on the local environment plays an important role in shaping the $R_{\rm e}$--$M_{\rm h}$ relation. Compared with the constant disruption model, the tidal field-based disruption prescription tends to enhance disruption in the inner parts of galaxies and increase the effective radius of GC systems (Fig.~\ref{fig:re_mh_log_with_constant}).

The kinematics of GC systems in our model is also consistent with observations. Most median systemic velocities, velocity dispersions, anisotropy parameter, pericenter/apocenter radii, and orbital actions of the modeled GC systems are consistent with the observational values of the MW (see, Table~\ref{tab:properties} and Figs.~\ref{fig:dispersion_r}, \ref{fig:v_sigma_mh_log}, and \ref{fig:beta_r_log}). However, the model predicts the median vertical velocity $v_z$ to be consistent with zero ($-5_{-10}^{+11}\kms$), whereas the MW GC system has systematically non-zero $v_{z,50}=14_{-7}^{+11}\kms$. Nevertheless, we still find some modeled systems to have even greater median $v_z$. The model also systematically overestimates the median $r_{\rm apo}$ for the MW GC system. This is likely because the MW GC system itself is more compact than the average MW mass system in our observational sample set; the model is more consistent with that average. The $v_{50}$--$M_{\rm h}$ and $\sigma_{50}$--$M_{\rm h}$ relations reveal that the systemic velocities of model GC systems are largely independent of host galaxy mass, whereas the velocity dispersions grow significantly as $M_{\rm h}$ increases (Fig.~\ref{fig:v_sigma_mh_log}). We also notice that the MW mass GC systems have positive anisotropy parameters growing from $\beta\simeq0$ to $0.5$ at $R=1-100$ kpc, indicating that the GC motions are more radially biased in outer parts of galaxies (Fig.~\ref{fig:beta_r_log}).

However, the GC mass function in our model peaks at lower mass compared with the MW GC system (Fig.~\ref{fig:mass_function}). This is possibly because the tidal disruption prescription is still not accounting for all relevant processes, including gravitational tidal shocks. 

By using galaxy merger trees from the adopted simulation, we can clearly identify the origins of GCs: \textit{in-situ} GCs form in the main progenitor branch of a given galaxy, whereas \textit{ex-situ} GCs form in satellite galaxies and later accrete onto the central galaxy. GCs with the \textit{in-situ} origin are systematically more concentrated towards the center, while \textit{ex-situ} GCs are found at larger radii out to $100$ kpc (Figs.~\ref{fig:Sigma_r_log_origins} and \ref{fig:re_mh_log_in_and_ex}). The \textit{in-situ} GCs are significantly more metal-rich than the \textit{ex-situ} ones because of the mass-metallicity relation for their host galaxies. The decreasing abundance of \textit{in-situ} GCs with radius leads to the metallicity gradient of the whole GC system (Fig.~\ref{fig:feh_r_log_origins}).

Our model also predicts notable differences between the kinematics of \textit{in-situ} and \textit{ex-situ} GCs. While the 3D velocity dispersion of both components decreases with radius, \textit{ex-situ} GCs have $\sim 40\kms$ higher velocity dispersion compared with the \textit{in-situ} counterparts at the same radius (Fig.~\ref{fig:dispersion_r_origins}). This higher dispersion is consistent with the migration nature of \textit{ex-situ} GCs, as they are brought into the main progenitor galaxy via accretion and mergers that can strongly perturb their original kinematics. Observationally, it is more applicable to compare dispersions of GCs split by the $\rm (B-I)$ color index: blue GCs have $\sim 40\kms$ higher velocity dispersion than the red counterparts, in agreement with observations of giant elliptical galaxies.

The \textit{in-situ} and \textit{ex-situ} GCs have systematically different median apocenter/pericenter radii and orbital actions (Table~\ref{tab:properties}). The deviation is especially notable for the median apocenter radius $r_{\rm apo}$, which is $3.5$ times greater for the \textit{ex-situ} GCs. Among the three orbital actions, the difference between the two GC populations is the largest for $J_R$ and the smallest for $|J_\phi|$, indicating that $J_R$ can be a useful parameter to distinguish GCs from different progenitors. However, note that we present here only the median values. Individual galaxies have a wide spread of the action variables, which makes it challenging to distinguish the progenitors of GCs. We will present more detailed investigation of the MW assembly history using GCs in a follow-up work.

\section*{Acknowledgements}
We are thankful for the insightful discussions with Hui Li, Xi Meng, Gillen Brown, Nick Choksi, Marta Reina-Campos, Monica Valluri, and Eric Bell. This research was mainly conducted with the \textsc{python} programming language, employing the following packages: \textsc{numpy} \citep{harris_array_2020}, \textsc{matplotlib} \citep{hunter_matplotlib_2007}, \textsc{scipy} \citep{virtanen_scipy_2020}, \textsc{agama} \citep{vasiliev_agama_2019}, and \textsc{illustris\_python} \citep{nelson_illustristng_2021}. The numerical analysis of this work was partially run on the Harvard Odyssey clusters with the account provided by Mark Vogelsberger and Lars Hernquist. OG and YC were supported in part by the U.S. National Science Foundation through grant 1909063. 

\section*{Data Availability}

The data that support the findings of this study are available from the corresponding author, upon reasonable request. The TNG50 simulation data are publicly available at \url{www.tng-project.org/data}.


\bibliographystyle{mnras}
\bibliography{GC-model-references} 


\appendix

\section{Accuracy of approximating the tidal tensor}
\label{sec:accuracy_of_approximating_tidal_disruption}

As mentioned in Sec.~\ref{sec:cluster_evolution}, we approximate the rate of tidal disruption via two methods based on the tidal tensor and local mass density, denoted as $\Omega_{\rm tid}=\Omega_{\lambda}$ and $\Omega_{\rm tid}=\Omega_{\rho}$, respectively. In this appendix, we describe tests of how accurately we can calculate the tidal tensor for the TNG50 simulation.

\subsection{Numerical test}

To numerically calculate the tidal tensor, we need to construct a $3\times3\times3$ grid with a side length of $d$ centered on a GC particle. To obtain an order of magnitude estimate for $d$, we refer to the resolution scale of the TNG50 simulation: for MW mass galaxies the median size of gas cells is around $0.1\ {\rm kpc}$, and the gravitational softening length of collisionless particles is $0.288\ {\rm kpc}$ \citep{pillepich_first_2019}. A too-large $d$ tends to smear out potential fluctuations and underestimate the strength of the tidal field, while a too-small $d$ differentiates the potential at a scale that is not numerically resolved, leading to unreliable results. Therefore we expect that an appropriate $d$ should have a similar value to the force resolution scale, varying between 0.1 and 1 kpc.

\begin{figure}
 	\includegraphics[width=\linewidth]{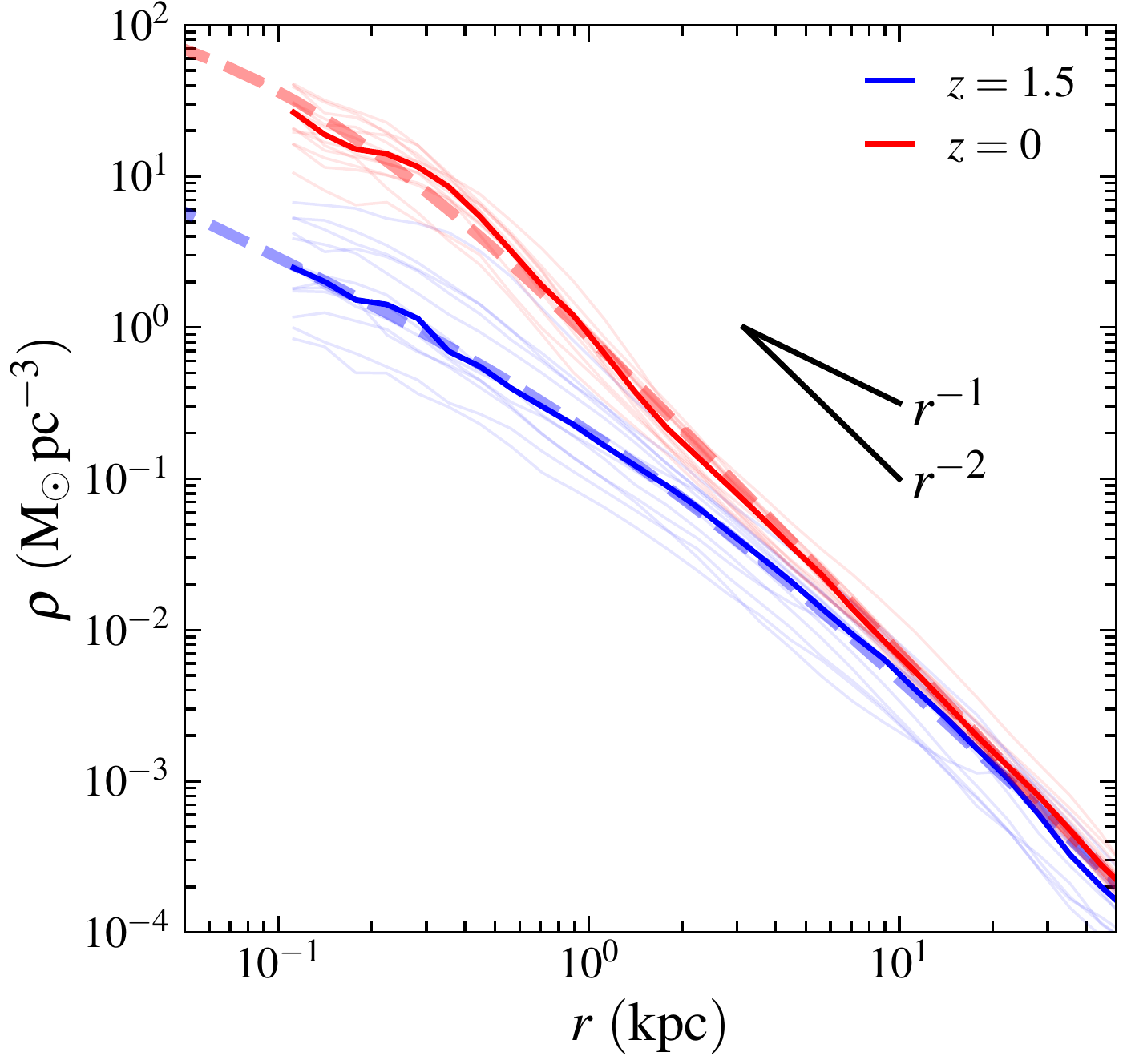}
 	\vspace{-4mm}
     \caption{Total mass density profiles of $13$ MW mass galaxies in TNG50 at $z=1.5$ (thin blue lines) and $z=0$ (thin red lines). The median density profiles are shown as thick solid lines. We also plot the double power-law fits at both epochs as thick dashed curves.}
     \label{fig:rho_r_tng}
\end{figure}

To assess the performance of our method with different $d$ quantitatively, we introduce the following test. First, we fit the mass distribution of the TNG50 galaxies used in our model with a parametric density profile, for which we can calculate the tidal field analytically. We take a double power-law profile as sufficiently general to describe most galaxy profiles:
\begin{equation}
	\rho(r) = \rho_0\left(
	\frac{r}{r_{\rm s}}\right)^{-\alpha}\left(1+
	\frac{r}{r_{\rm s}}\right)^{-(\beta-\alpha)},
\end{equation}
where $\alpha$ measures the inner slope of profile for $r\ll r_{\rm s}$, while $\beta$ measures the outer slope for $r\gg r_{\rm s}$. We include all matter components (stars, gas, and dark matter) in the fit because they all contribute to the tidal field. For MW mass galaxies at $z=0$, the fitting result yields $\alpha=0$, $\beta=2.2$, and $r_{\rm s}=0.1\ {\rm kpc}$. The very small $r_{\rm s}$ indicates that the mass distribution roughly follows an isothermal profile over a wide range of radii down to $r\simeq0.1\ {\rm kpc}$. We also perform the fitting for the same galaxies at $z=1.5$ to better compare with \citet{meng_tidal_2022}, who analyzed in detail the tidal disruption rate for star clusters at $z\simeq 1.5$ in mesh-based high-resolution simulations. For these high-redshift galaxies we find $\alpha=1$, $\beta=2$, and $r_{\rm s}=1.5\ {\rm kpc}$, which describes a combination of an isothermal profile at large radii and a flatter inner core. These profiles are shown in Fig.~\ref{fig:rho_r_tng}.

Then we construct a mock particle realization of the density profile, with the same mass of $2.7\times10^{5}\,\Msun$ as the average particle mass in TNG50. For our fitting parameters, there are $2.7$ million particles within $100\ {\rm kpc}$ at $z=0$, and $2.4$ million within $100\ {\rm kpc}$ at $z=1.5$.

\begin{figure*}
 	\includegraphics[width=0.9\linewidth]{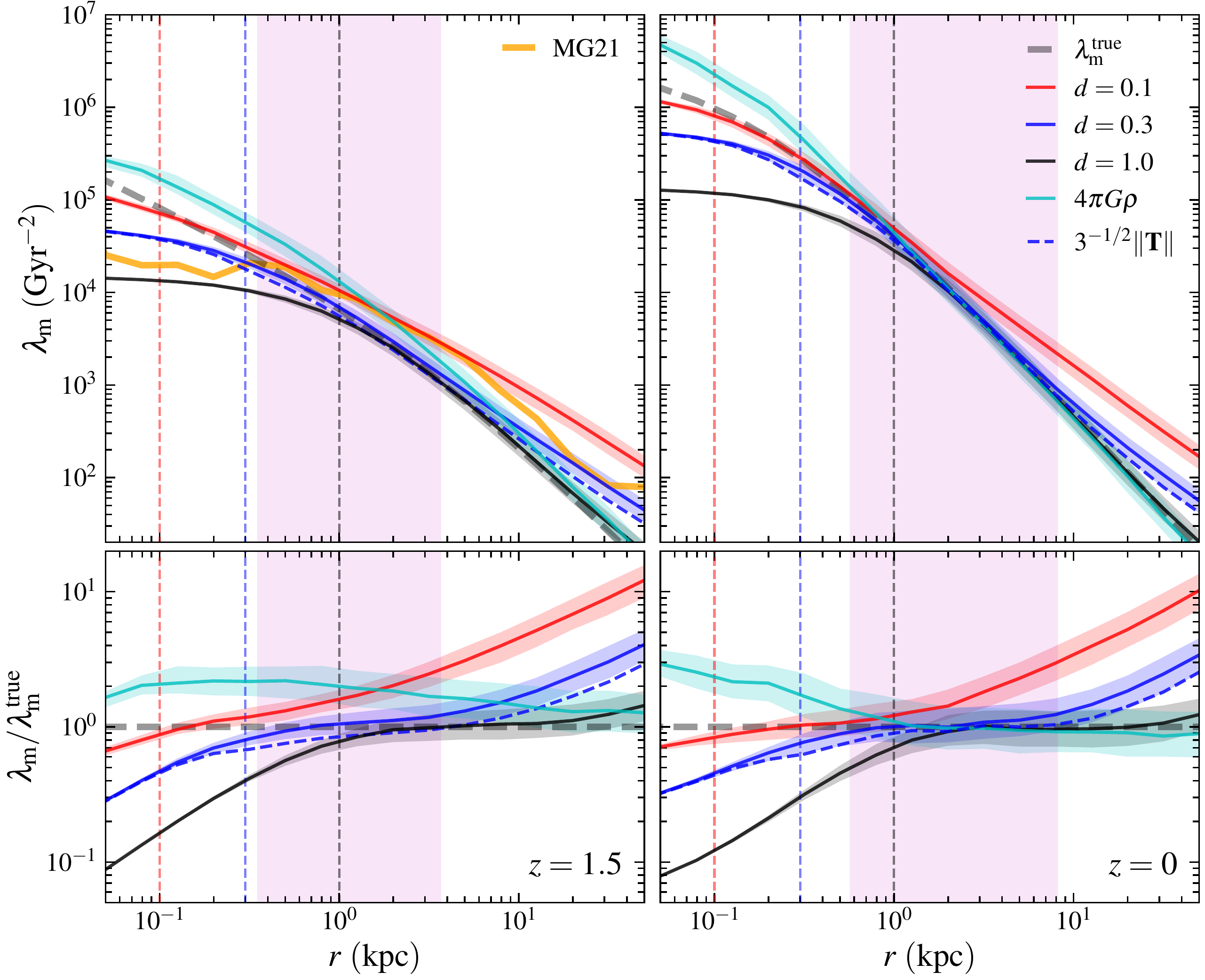}
 	\vspace{0mm}
     \caption{Radial profiles of approximated $\lambda_{\rm m}$ (\textit{top panels}) and the deviation from analytical results, $\lambda_{\rm m}/\lambda_{\rm m}^{\rm true}$ (\textit{bottom panels}), calculated with $d=0.1$ (red), $0.3$ (blue), and $1.0\ {\rm kpc}$ (black) for two mock galaxies, which are generated with double power-law profiles to represent MW mass galaxies in TNG50 at $z=1.5$ (left) and $z=0$ (right). We plot three vertical dashed lines to represent the location of each $d$. Approximated density profile, rescaled by a factor of $4\pi G$, is shown as the cyan curve. The $16-84$th percentiles of approximated values are shown as shaded regions. We also plot a gray dashed curve to show the analytical result, $\lambda_{\rm m}^{\rm true}$. To compare with higher resolution simulations, we show the $\lambda_{\rm m}$-$r$ relation for MW galaxies at $z\simeq1.5$ from \citet{meng_tidal_2022} orange solid curves. The pink shaded regions represent the $16-84$th percentiles of the radial distribution of all modeled GCs (both surviving and disrupted) at $z=1.5$ and $z=0$.}
     \label{fig:lambda_r_log}
\end{figure*}

Finally, we apply the same method we have described in Sec.~\ref{sec:cluster_evolution} to particles in the mock galaxy and calculate the tidal tensor with $d=0.1$, $0.3$, and $1.0\ {\rm kpc}$. We plot the largest eigenvalue of the tidal tensor as a function of the galactocentric radius in Fig.~\ref{fig:lambda_r_log}. By comparing the calculated value to the analytical value $\lambda_{\rm m}^{\rm true}$, we find that different $d$ work best at different radii. All three cases underestimate $\lambda_{\rm m}$ when $r\lesssim d$ since the details of the tidal field are smeared out on small scales. Moreover, the approximation also fails at large radii, because the average separation between particles becomes too large at $r \gtrsim 10$~kpc. The tidal field gets under-resolved when the average separation is comparable to or even larger than $d$. In our model, majority of GCs particles (regardless of whether they survive or disrupt) are located at $r=0.3-3\ {\rm kpc}$ at $z=1.5$ and $r=0.5-8\ {\rm kpc}$ at $z=0$. We note that the tidal tensor in both ranges can be approximated to within $0.1\ {\rm dex}$ by adopting $d=0.3\ {\rm kpc}$. This conclusion is confirmed by \citet{meng_tidal_2022}, who compared the time-averaged $P$ parameter (which can be considered as an indicator of tidal strength, see their Eq.~(4) and (6)) of their simulations and this work with $d=0.3\ {\rm kpc}$. In their Fig.~12, they plotted $P$ of each cluster as a function of the host halo mass at formation. Their results are in good agreement with ours, both showing a decreasing trend with large scatters. This consistency supports that setting $d=0.3\ {\rm kpc}$ can to a large extent approximate the tidal field in GC disruption.

\citet{meng_tidal_2022} performed a similar analysis of the $\lambda_{\rm m}$ profile. They calculated the tidal tensor of star clusters from a suite of high-resolution cosmological simulations of \citet{li_star_2017}, which are run with adaptive mesh refinement with the finest refinement level reaching $30\ {\rm pc}$ at $z=1.5$. Therefore, they can resolve the tidal field at a scale of $30\ {\rm pc}$ in the densest region of galaxies. We note that $\lambda_{\rm m}$ from \citet{meng_tidal_2022} have lower value at $r\lesssim 0.3\ {\rm kpc}$, since the feedback in their work is stronger than TNG50, leading to flatter density profiles and therefore weaker tidal field in the galaxy center. Nevertheless, their $\lambda_{\rm m}$ are greater than that of the mock galaxy by a factor of $\sim3$ at $r\gtrsim 1\ {\rm kpc}$. This is because our isotropic density profile is too smooth to correctly show the asymmetric density fluctuations revealed by higher-resolution simulations. Similarly, since the gravitational softening length of TNG50 ($0.288\ {\rm kpc}$) is much larger than the typical size of a GC, we tend to underestimate $\lambda_{\rm m}$ as the small-scale density fluctuations in TNG50 are also over-smoothed. The interplay of all factors mentioned above motivates us to apply the parameter $\kappa_\lambda$ to correct the calculation of the tidal tensor.

As described in Sec.~\ref{sec:cluster_evolution}, we also employ $4\pi G\rho$ as an estimate of $\lambda_{\rm m}$. Fig.~\ref{fig:lambda_r_log} shows that we can well estimate $\lambda_{\rm m}$ at large radii where both the $z=1.5$ and $z=0$ cases are nearly isothermal. However, at smaller radii, where the density profiles are flatter, $4\pi G\rho$ tend to overestimate $\lambda_{\rm m}$ by a factor of $2$ to $3$. It is therefore reasonable to expect smaller $\kappa_{\rho}$ than $\kappa_{\lambda}$ if we want the two models to produce similar numbers of GCs. However, this may not be the case since we do not only take into account the number of GCs when performing model calibration, see Sec.~\ref{sec:merit_function}.

\subsection{Analytical approximation}

\begin{figure}
 	\includegraphics[width=\linewidth]{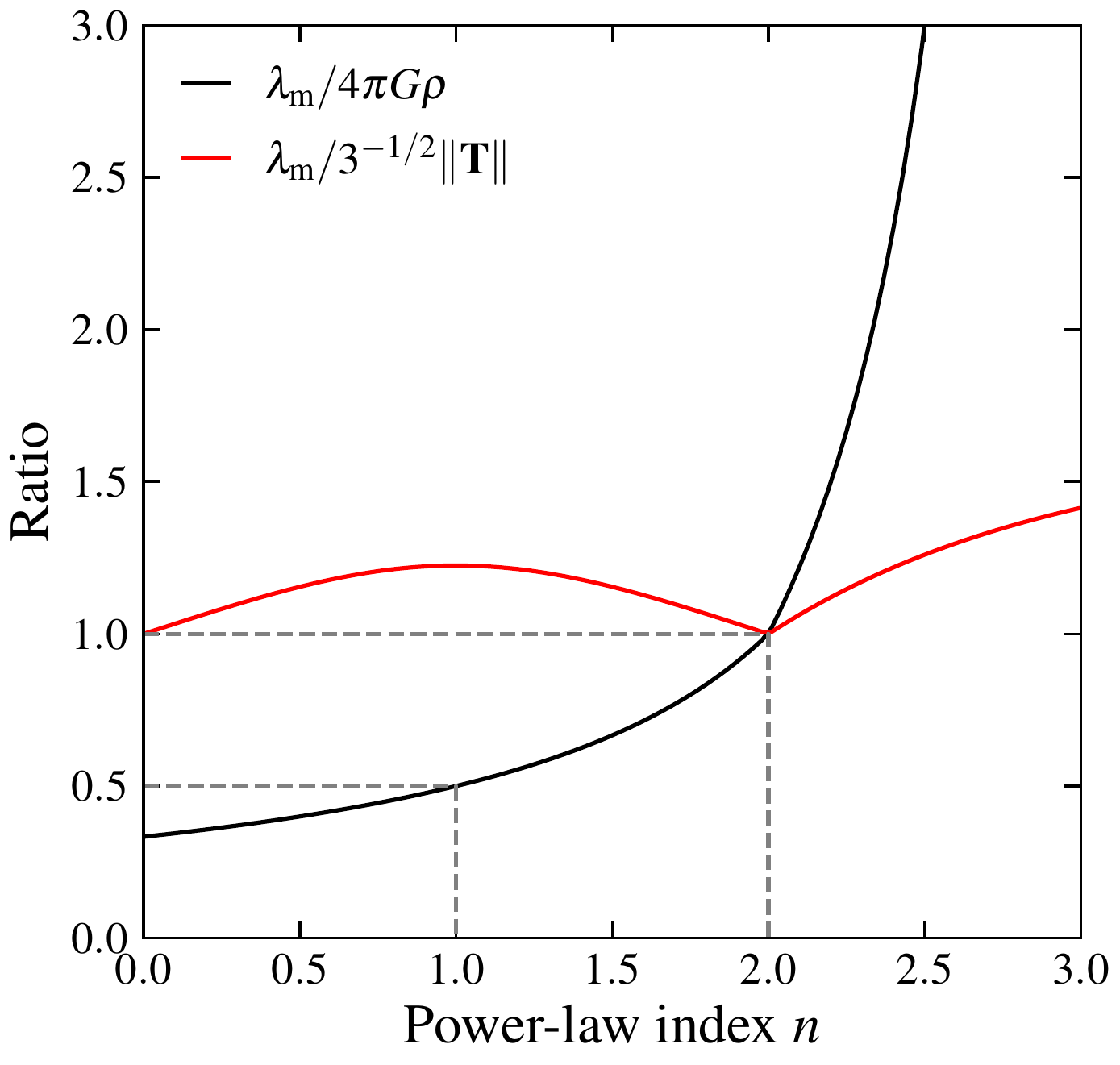}
 	\vspace{-4mm}
     \caption{Analytical $\lambda_{\rm m}/4\pi G\rho$ (black line) and $\lambda_{\rm m}/3^{-1/2}\|{\bf T}\|$ (red line) for power-law density profiles, $\rho\propto r^{-n}$. Dashed lines indicate special cases of $n=1$ and $n=2$.}
     \label{fig:rho_over_lambda_n}
\end{figure}

To better understand the deviation of $\lambdam$ from $4\pi G\rho$, we perform an analytical analysis as follows. Usually, the tidal tensor is defined in the Cartesian coordinate system, see Eq.~(\ref{eq:tidal_tensor}). Since the aforementioned scenarios are spherically symmetric, it is convenient to analyze the tidal tensor in spherical coordinate systems. By defining $\mu_\theta=\cos\theta$, $\nu_\theta=\sin\theta$, $\mu_\phi=\cos\phi$, and $\nu_\phi=\sin\phi$, we can write the coordinate transformation as
\begin{equation}
	\left\{ 
	\begin{array}{lll}
		x=r\mu_\theta\mu_\phi, \\
		y=r\mu_\theta\nu_\phi, \\
		z=r\nu_\theta. \\
	\end{array}
	\right.
\end{equation}
Therefore, the Jacobian matrix is given by
\begin{equation}
	{\bf J}\equiv\frac{\partial\mathbfit{x}}{\partial\mathbfit{x}'}=\left( 
	\begin{array}{ccc} 
		\mu_\theta\mu_\phi & -r\nu_\theta\mu_\phi & -r\mu_\theta\nu_\phi \\
		\mu_\theta\mu_\phi & -r\nu_\theta\nu_\phi & r\mu_\theta\mu_\phi \\
		\nu_\theta & r\mu_\theta &
	\end{array} 
	\right),
\end{equation}
where $\mathbfit{x}=(x,y,z)$ and $\mathbfit{x}'=(r,\theta,\phi)$. According to the chain rule for partial derivatives, we can write the Cartesian partial derivatives in spherical coordinates as
\begin{equation}
	\frac{\partial}{\partial\mathbfit{x}} = ({\bf J}^{-1})^{\rm T}\frac{\partial}{\partial\mathbfit{x}'},
\end{equation}
where ${\bf J}^{-1}$ is the inverse of ${\bf J}$:
\begin{equation}
	{\bf J}^{-1}=\left( 
	\begin{array}{ccc} 
		\mu_\theta\mu_\phi & \mu_\theta\nu_\phi & \nu_\theta \\
		-\nu_\theta\mu_\phi/r & -\nu_\theta\nu_\phi/r & \mu_\theta/r \\
		-\nu_\phi/r\mu_\theta & \mu_\phi/r\mu_\theta &
	\end{array} 
	\right).
\end{equation}
Similarly, the second Cartesian partial derivatives are
\begin{equation}
	\frac{\partial}{\partial\mathbfit{x}}\left(\frac{\partial}{\partial\mathbfit{x}}\right)^{\rm T}
	= ({\bf J}^{-1})^{\rm T}\frac{\partial}{\partial\mathbfit{x}'}\left[({\bf J}^{-1})^{\rm T}\frac{\partial}{\partial\mathbfit{x}'}\right]^{\rm T}.
\end{equation}
Therefore, we can rewrite Eq.~(\ref{eq:tidal_tensor}) as
\begin{equation}
	{\bf T} = -\frac{\partial}{\partial\mathbfit{x}}\left(\frac{\partial\Phi}{\partial\mathbfit{x}}\right)^{\rm T} = -({\bf J}^{-1})^{\rm T}\frac{\partial}{\partial\mathbfit{x}'}\left[({\bf J}^{-1})^{\rm T}\frac{\partial\Phi}{\partial\mathbfit{x}'}\right]^{\rm T}.
\end{equation}
Recall that the potential is spherically symmetric, i.e., $\partial\Phi/\partial\theta=\partial\Phi/\partial\phi=0$, and the above equation simplifies to
\begin{equation}
	{\bf T} = - \left({\bf A}\cdot\frac{d^2\Phi}{d r^2} + {\bf B}\cdot\frac{1}{r}\frac{d\Phi}{d r}\right),
\end{equation}
where
\begin{equation}
	{\bf A} = \left( 
	\begin{array}{ccc} 
		\mu_\theta^2\mu_\phi^2 & \mu_\theta^2\mu_\phi\nu_\phi & \mu_\theta\nu_\theta\mu_\phi \\
		\mu_\theta^2\mu_\phi\nu_\phi & \mu_\theta^2\nu_\phi^2 & \mu_\theta\nu_\theta\nu_\phi \\
		\mu_\theta\nu_\theta\mu_\phi & \mu_\theta\nu_\theta\nu_\phi & \nu_\theta^2
	\end{array} 
	\right),
\end{equation}
and
\begin{equation}
	{\bf B} = \left( 
	\begin{array}{ccc} 
		(\mu_\theta^2+1)\nu_\phi^2 & -\mu_\theta^2\mu_\phi\nu_\phi & -\mu_\theta\nu_\theta\mu_\phi \\
		-\mu_\theta^2\mu_\phi\nu_\phi & \mu_\phi^2+\nu_\theta^2\nu_\phi^2 & -\mu_\theta\nu_\theta\nu_\phi \\
		-\mu_\theta\nu_\theta\mu_\phi & -\mu_\theta\nu_\theta\nu_\phi & \mu_\theta^2
	\end{array} 
	\right).
\end{equation}
The three eigenvalues of ${\bf T}$ are
\begin{equation}
	\lambda_1=-\frac{d^2\Phi}{d r^2}, \quad \lambda_2=\lambda_3=-\frac{1}{r}\frac{d\Phi}{d r}.
\end{equation}
Therefore, we obtain
\begin{equation}
	\lambda_{\rm m}(r) \equiv \max|\lambda_i| = \max\left(\left|\frac{d^2\Phi}{d r^2}\right|,\ \left|\frac{1}{r}\frac{d\Phi}{d r}\right|\right).
\end{equation}
For a spherically symmetric density distribution,
\begin{equation}
	\Phi(r)=-4\pi G\left[\frac{1}{r}\int_0^r\rho(r')r'^2dr'+\int_r^\infty\rho(r')r'dr'\right],
\end{equation}
we get
\begin{equation}
	\lambda_{\rm m}(r) = \max\left(\left|4\pi G\rho-\frac{2GM(<r)}{r^3}\right|,\ \frac{GM(<r)}{r^3}\right),
\end{equation}
where $M(<r)$ denotes the mass enclosed within radius $r$. For a special case of power-law density profiles, $\rho\propto r^{-n}$,
\begin{equation}
	\lambda_{\rm m} = 4\pi G\rho \; \max\left(\left|\frac{1-n}{3-n}\right|,\ \left|\frac{1}{3-n}\right|\right).
\end{equation}
We show this relation in Fig.~\ref{fig:rho_over_lambda_n}. For a special case of the isothermal profile with $n=2$, the maximum eigenvalue is particularly simple: $\lambda_{\rm m} = 4\pi G\rho$. For flatter profiles, $4\pi G\rho$ tends to overestimate $\lambda_{\rm m}$ by up to a factor of $2$ at $n=1$, in agreement with the results shown in Fig.~\ref{fig:lambda_r_log}. For a simulated galaxy with $n$ ranging from $1$ to $2$, we therefore expect $\kappa_\lambda/\kappa_\rho \lesssim\sqrt{2}$, which agrees with the calibration results in Sec.~\ref{sec:parameter_selection}.

Another proxy for $\Omega_{\rm tid}$ is the Frobenius norm of the tidal tensor, 
\begin{equation}
	\|{\bf T}\|\equiv\left(\sum_i\sum_j |T_{ij}|^2\right)^{1/2}=\left(\sum_i|\lambda_i|^2\right)^{1/2}.
\end{equation}
The last equality holds because ${\bf T}$ can be diagonalized. For power-law density profiles, we get
\begin{equation}
	\|{\bf T}\|=4\pi G\rho\left(\frac{n^2-2n+3}{3-n}\right)^{1/2}.
\end{equation}
In Fig.~\ref{fig:rho_over_lambda_n}, we show the ratio between $\lambda_{\rm m}$ and $\|{\bf T}\|$ rescaled by a factor of $3^{-1/2}$. The two identities only differ by less than $50\%$ for $n=0- 3$, meaning that we can also employ the Frobenius norm as a proxy for $\Omega_{\rm tid}$ with similar results as the $\Omega_{\rm tid}=\Omega_{\lambda}$ case.

\section{Test of different model parameters}
\label{sec:test_of_different_model_parameters}

\begin{figure}
	\includegraphics[width=\linewidth]{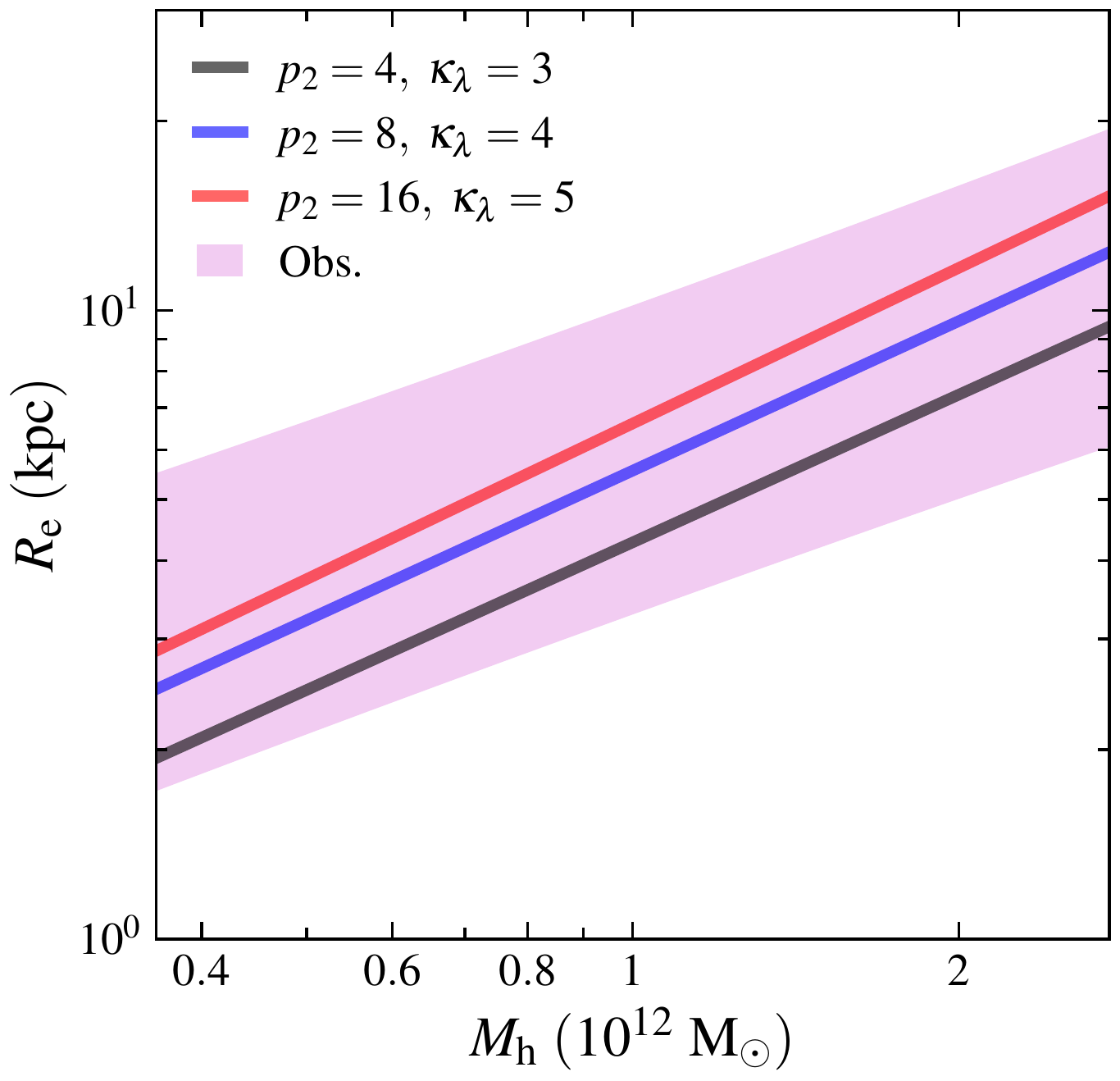}
	\vspace{-3mm}
    \caption{$R_{\rm e}$--$\Mh$ relations for the $\Omega_{\rm tid}=\Omega_\lambda$ cases with $(p_2,\kappa_\lambda)=(4,3)$ (black), $(8,4)$ (blue), and $(16,5)$ (red). Other parameters are as in Fig.~\ref{fig:re_mh_log}.}
    \label{fig:re_mh_log_kappa}
\end{figure}

In this work, we select model parameters $(p_2,p_3,\kappa)$ by minimizing the merit function $\cal M$. However, a large region in the $(p_2,p_3,\kappa)$ space results in similarly small $\cal M$. This is due to the $p_2$--$\kappa$ degeneracy in our model. As mentioned in Sec.~\ref{sec:cluster_formation}, $p_2$ characterizes the total mass of GCs in each cluster formation event, i.e., the strength of GC formation. On the other hand, $\kappa$ quantifies the strength of tidal disruption, see Sec.~\ref{sec:cluster_evolution}. Therefore, the effect of increasing $p_2$ can be largely canceled by increasing $\kappa$. As a consequence, a wide range $p_2$ can lead to similarly small $\cal M$ by adjusting $\kappa$ accordingly. To test this degeneracy, we fix $p_3$ to $0.5\ {\rm Gyr^{-1}}$ and try different $(p_2,\kappa_\lambda)$ configurations of $(4,3)$, $(8,4)$, and $(16,5)$ for the $\Omega_{\rm tid}=\Omega_\lambda$ case. The GC kinematics vary little with different $(p_2,\kappa_\lambda)$, while the radial distributions of the three cases are somewhat different, as illustrated in Fig.~\ref{fig:re_mh_log_kappa}. We note that the $R_{\rm e}$--$\Mh$ relations have similar power-law indices in all three cases, whereas the larger $\kappa$ prescriptions tend to predict larger effective radii $R_{\rm e}$. This is likely because higher fractions of inner GCs are disrupted in the higher $\kappa$ cases.

Even though the three cases have vastly different $p_2$ varying from $4$ to $16$, all the predicted $R_{\rm e}$--$\Mh$ relations match the observations within the 1-$\sigma$ confidence level, indicating that all the configurations can serve as an appropriate choice for the model. Without loss of generality, we select the configuration $(p_2,p_3,\kappa_\lambda)=(8,0.5\ {\rm Gyr^{-1}},4)$ for the $\Omega_{\rm tid}=\Omega_\lambda$ model throughout the paper. In the $\Omega_{\rm tid}=\Omega_\rho$ case, we note a similar degeneracy and choose $(p_2,p_3,\kappa_\rho)=(8,0.5\ {\rm Gyr^{-1}},5)$. 

An alternative prescription of tidal field takes into account fictitious forces: centrifugal, Euler, and Coriolis \citep[see,][]{renaud_evolution_2011}. Following \citet{pfeffer_e-mosaics_2018}, we can approximate the effective tidal strength as $\lambda_{\rm 1,e} \approx \lambda_1 - 0.5(\lambda_2+\lambda_3)$, where $\lambda_{\rm 1}$ is the maximum eigenvalue of the tidal tensor (without using absolute values). Similarly to the disruption models in this work, we can approximate the tidal angular frequency by $\Omega_{\rm tid}^2=\kappa_{\rm e}\lambda_{\rm 1,e}$. After calibration, we find the best-fit parameters to be $(p_2,p_3,\kappa_{\rm e})=(8,0.5\ {\rm Gyr^{-1}},2)$. Since the GC properties predicted by the $\Omega_{\rm tid}^2=\kappa_{\rm e}\lambda_{\rm 1,e}$ model are consistent with the $\Omega_{\rm tid}=\Omega_{\lambda}$ case, we do not present the results of this model for brevity.

\bsp	
\label{lastpage}
\end{document}